\pdfoutput=1

\documentclass[a4paper,11pt]{article} 

\usepackage{atlasphysics}
\usepackage[displaymath]{lineno}
\usepackage{jheppub} 

\usepackage{graphicx} 
\usepackage{subfigure}
\usepackage{amsmath}
\usepackage{amssymb}
\usepackage{multirow}
\usepackage{lscape}
\usepackage{lineno}
\usepackage{rotating}
\usepackage{slashed}
\usepackage{url}
\usepackage{xspace}
\usepackage{booktabs}
\usepackage[section]{placeins}
\usepackage{flafter}

\usepackage{hyperref}
\hypersetup{
  colorlinks=true,
  linkcolor=red,
  citecolor=red,
  urlcolor=blue
}

\input{tools.tex}

\newcommand{\ExpLimitResolved}{639~{\rm GeV}}
\newcommand{\ObsLimitResolved}{666~{\rm GeV}}
\newcommand{\ExpLimitBoosted}{269~{\rm GeV}}
\newcommand{\ObsLimitBoosted}{255~{\rm GeV}}
\newcommand{\intlumi}{$4.6 \pm 0.2$\invfb}

\graphicspath{{figures/}}
\usepackage{preprintcover}  
\PreprintCoverPaperTitle{Search for pair production of massive particles decaying into three quarks with the ATLAS detector in \sqsseven \pp collisions at the LHC}
\PreprintIdNumber{CERN-PH-EP-2012-281}
\PreprintCoverAbstract{A search is conducted for hadronic three-body decays of a new 
  massive coloured particle in \sqsseven \pp collisions at the LHC using an integrated 
  luminosity of 4.6\invfb collected by the ATLAS detector. Supersymmetric gluino 
  pair production in the context of a model with $R$-parity violation is used as 
  a benchmark scenario. The analysis is divided into two search channels, each 
  optimised separately for their sensitivity to high-mass and low-mass gluino production. 
  The first search channel uses a stringent selection on the transverse 
  momentum of the six leading jets and is performed as a counting experiment. 
  The second search channel focuses on low-mass gluinos produced with a large
  boost. Large-radius jets are selected and the invariant mass of each of 
  the two leading jets is used as a discriminant between the signal and the 
  background. The results are found to be consistent with 
  Standard Model expectations and limits are set on the allowed gluino mass.}
\PreprintJournalName{JHEP}

\title{Search for pair production of massive particles decaying into three quarks with the
       ATLAS detector in \sqsseven \pp collisions at the LHC}

\author{The ATLAS Collaboration}
                             
\abstract{A search is conducted for hadronic three-body decays of a new 
  massive coloured particle in \sqsseven \pp collisions at the LHC using an integrated 
  luminosity of 4.6\invfb collected by the ATLAS detector. Supersymmetric gluino 
  pair production in the context of a model with $R$-parity violation is used as 
  a benchmark scenario. The analysis is divided into two search channels, each 
  optimised separately for their sensitivity to high-mass and low-mass gluino production. 
  The first search channel uses a stringent selection on the transverse 
  momentum of the six leading jets and is performed as a counting experiment. 
  The second search channel focuses on low-mass gluinos produced with a large
  boost. Large-radius jets are selected and the invariant mass of each of 
  the two leading jets is used as a discriminant between the signal and the 
  background. The results are found to be consistent with 
  Standard Model expectations and limits are set on the allowed gluino mass.
}

\begin{document}

\maketitle
\flushbottom

\clearpage

\section{Introduction}
\label{sec:introduction}
Supersymmetry (\SUSY)~\cite{Miyazawa:1966,Ramond:1971gb,Golfand:1971iw,Neveu:1971rx,Neveu:1971iv,Gervais:1971ji,Volkov:1973ix,Wess:1973kz,Wess:1974tw} is a theoretical extension of the Standard Model (SM), where a new symmetry relates fermions and bosons. \SUSY has the potential to solve the hierarchy problem~\cite{Dimopoulos:1981zb,Witten:1981nf,Dine:1981za,Dimopoulos:1981au,Sakai:1981gr,Kaul:1981hi} and to provide a dark matter candidate~\cite{Goldberg:1983nd,Ellis:1983ew}. As a result of the latter possibility, most searches for \SUSY focus on scenarios such as the ``minimal'' supersymmetric extension of the Standard Model (MSSM) in which $R$-parity is conserved~\cite{SUSYZeroLep2011, SUSYOneLep2011, SUSYTwoLep2011, SUSYJetMult2011}. In R-parity-conserving (RPC) models, \SUSY particles must be produced in pairs and must decay in a sequence which ends with the lightest stable supersymmetric particle (LSP). However, with strong constraints now placed on standard RPC \SUSY scenarios by the LHC experiments, it is important to ensure a broad scope for the \SUSY search program.

In $R$-parity-violating (RPV) models, many of the constraints placed on the MSSM in terms of the allowed parameter space of the masses of the SUSY partners of the gluons and quarks, the gluinos (\gluino) and squarks (\squark), are relaxed. The reduced sensitivity of standard \SUSY searches to RPV scenarios is primarily due to the high missing transverse momentum requirements used in the event selection. These choices are motivated by the assumed presence of an undetected LSP and strongly reduce SM background contributions. For RPV SUSY, different approaches must be used depending on the targeted scenarios.

In this paper, a search is presented for fully hadronic final states involving massive particle decays to three jets. An RPV SUSY model in which pair produced gluinos each decay to three jets via an off-shell squark ($\tilde{g} \rightarrow q\tilde{q} \rightarrow qqq$\ with $m_{\tilde{q}} >> m_{\tilde{g}}$) is used as a benchmark physics model. Two complementary methods are used to distinguish the signal from the SM multijet background, both using \intlumi\ of data collected at \sqsseven. The first (resolved) analysis channel resolves all six jets in order to search for an excess in the jet multiplicity spectrum. Whereas the pair production of very massive gluinos tends to produce final states with six well-separated jets, event signatures from the low and intermediate mass range is considerably more difficult to identify. The second (boosted) analysis channel exploits the collimation of the decay products that is expected when the gluinos are boosted. Gluinos produced with a large momentum relative to their mass may therefore result in overlapping jets from each of the three quarks.  In this case, a large-radius jet algorithm is used to capture the three-body decay products in a single jet. The mass of such jets, as well as properties of their internal structure that are characteristic of the presence of a massive boosted object, provide discrimination against the SM multijet background. This approach not only serves as a cross-check of the resolved method, but also provides an orthogonal search channel with a nearly independent set of systematic uncertainties and represents the first such application of jet substructure techniques in the search for SUSY at the LHC.

Other searches for such final states have been conducted by the CDF~\cite{Aaltonen:2011sg} and the CMS~\cite{Chatrchyan:2011cj, CMSRPVGluino2011} collaborations. The CMS results use a nearly identical signal model to that considered here and report limits which restrict the allowed ranges of gluino masses to $144 < \mgluino < 200$\ GeV or $\mgluino>480$~GeV, using approximately 5\invfb of data at \sqsseven. 

This paper is organised as follows. \Secref{Detector} describes the ATLAS detector and the data samples used to conduct the search. \Secref{data-mc} describes the simulated samples used for the signal and background studies. \Secref{Resolved} and \secref{Boosted} present the details of the event selection, background estimations, and systematic uncertainties used in the resolved and boosted analysis techniques, respectively. The final combined results and exclusion limits on the RPV gluino model tested are shown in \secref{Results}.

\section{The ATLAS detector and data samples}
\label{sec:Detector}
The ATLAS detector~\cite{detPaper,PerfWithData2010} provides nearly full solid angle coverage around the collision point with an inner tracking system covering $|\eta|<2.5$\footnote{ATLAS uses a right-handed coordinate system with its origin at the nominal interaction point (IP) in the centre of the detector and the $z$-axis along the beam pipe. The $x$-axis points from the IP to the centre of the LHC ring, and the $y$-axis points upward. Cylindrical coordinates $(r,\phi)$ are used in the transverse plane, $\phi$ being the azimuthal angle around the beam pipe. The pseudorapidity is defined in terms of the polar angle $\theta$ as $\eta=-\ln\tan(\theta/2)$.},  electromagnetic and hadronic calorimeters covering $|\eta|<4.9$, and a muon spectrometer covering $|\eta|<2.7$. For this analysis the most relevant ATLAS subsystems are the barrel and end-cap calorimeters~\cite{LArReadiness, TileReadiness} and the trigger system~\cite{TriggerPerf2010}. 



The calorimeter comprises multiple subdetectors with several different designs, spanning the pseudorapidity range up to $|\eta|=4.9$. The measurements presented here are predominantly performed using data from the central calorimeters that consist of the Liquid Argon (LAr) barrel electromagnetic calorimeter ($|\eta|<1.475$) and the Tile hadronic calorimeter ($|\eta|<1.7$). Three additional calorimeter subsystems are located in the forward regions of the detector: the LAr electromagnetic end-cap calorimeters ($1.375<|\eta|<3.2$), the LAr hadronic end-cap calorimeter ($1.5<|\eta|<3.2$), and the forward calorimeter that features separate EM and hadronic compartments ($3.1<|\eta|<4.9$). As described below, jets are required to have $|\eta|<2.8$ such that they are fully contained within the barrel and end-cap calorimeter systems.
 
The precision and accuracy of energy measurements~\cite{jespaper2010} made by the calorimeter system are integral to this analysis. Electrons and muons produced in test-beams are used to establish the baseline electromagnetic (EM) energy scale of the LAr and Tile calorimeters~\cite{ctb2004electronseoverp, ctb2004electrons, LArTB02uniformity, LArTB02linearity, LArTB02muons, Tile2002}. The response to pions was also measured using test-beams and is used to validate the detector simulation model for both the EM and hadronic calorimeters~\cite{Tile2002, Tile2002pionproton, CTB2004topology, CTB04pion, CTB2004vlepion, EndcapTBelectronPion2002, Pinfold:2008zzb, Kiryunin:2006cm}.  Further \insitu\ measurements using cosmic-ray muons are used to validate the hadronic calorimeter's energy scale in the experimental hall~\cite{TileReadiness}. The invariant mass of the $Z$ boson in $Z\rightarrow ee$ events measured \insitu\ is used to adjust the calibration for the EM calorimeters~\cite{Atlaselectronpaper}. 

The jets used for this analysis are found and reconstructed using the \antikt\ algorithm~\cite{Cacciari200657,Cacciari:2008gp} with a radius parameter $R = 0.4$. To construct the input to the calorimeter jet finding, a local cluster weighting calibration method~\cite{TopoClusters} first clusters together topologically connected calorimeter cells and classifies these so-called ``\topos'' as either electromagnetic or hadronic. Based on this classification, energy corrections are applied that are derived from single-pion simulations. Dedicated hadronic corrections are determined for the effects of non-compensation, signal losses due to noise-suppression threshold effects, and energy lost in non-instrumented regions. The final jet energy calibration is derived as a correction relating the calorimeter's response to the true jet energy based upon simulation~\cite{jespaper2010}. 

Dedicated trigger and data acquisition systems are responsible for the online event selection, which is performed in three stages: Level 1, Level 2, and the Event Filter. The measurements presented in this paper use single-jet and multijet triggers which, for the analysis selections used, are more than 99\% efficient. In particular, the multijet triggers implemented at the Event Filter level have access to the full detector granularity, which allows selection of multijet events with high efficiency.

Data from the entire 2011 ATLAS data-taking period is used. All data are required to have met baseline quality criteria and were taken during periods in which the detector operated without problems. Data quality criteria reject events with significant contamination from detector noise or issues in the read-out and are based upon individual assessments for each subdetector. After removing these events the remaining data corresponds to approximately \intlumi\ of integrated luminosity~\cite{Aad:2011dr,ATLAS-CONF-2011-116}. Multiple proton-proton (\pp) collisions, or pile-up, result in several reconstructed primary vertices per event. The hard scattering vertex is selected by choosing the vertex with the maximum sum of the squared track transverse momenta, $\sum(\pttrk)^{2}$, from vertices that have at least two tracks with $\pttrk > 0.4$~GeV.

\section{Monte Carlo samples}
\label{sec:data-mc}

Monte Carlo (MC) events are used to model the signal efficiency, to optimise the event selection requirements and to aid in the description of the SM backgrounds. Signal MC samples, consisting of pair-produced gluinos, each decaying to three quarks via an off-shell squark, are generated using \Madgraph 5 version 1.3.33~\cite{Alwall:2007st, Alwall:2011uj} with the RPVMSSM~\cite{Fuks:2012im} model used to perform the matrix element calculations. In this paper we choose to probe couplings that will produce a fully-hadronic final state. Therefore, the parameters that allow gluinos to decay into top quarks are set to be zero. We further set the couplings to values such that the gluinos decay with a negligible lifetime. The resulting parton-level events are interfaced to \Pythia 8.160~\cite{pythia8} for showering, hadronisation, and underlying event (UE) simulation. Signal cross-sections are calculated to next-to-leading order (\NLO) precision in the strong coupling constant, adding the resummation of soft gluon emission at next-to-leading-logarithmic accuracy (\NLL)~\cite{Beenakker:1996ch, Kulesza:2008jb, Kulesza:2009kq, Beenakker:2009ha, Beenakker:2011fu} accuracy. The nominal cross-section and the uncertainty are taken from an envelope of cross-section predictions using different parton distribution function (PDF) sets and factorisation and renormalisation scales~\cite{Kramer:2012bx}. The following mass points are used to evaluate the sensitivity: $\mgluino=100, 200, 300, 400, 500, 600$ and 800~GeV. In this paper, all superpartners except for the gluinos are set to have a mass of 5~TeV, corresponding to a model with decoupled squarks. In models with squark masses that are much smaller than this, the kinematics of the signal depend on the properties of the squarks in the cascade decays. It should be noted that some reinterpretation would be needed to apply the results of this paper to such cases.


%
Dijet and multijet events are simulated in order to study the background SM contributions and background estimation techniques. Both a leading-order (LO) matrix element (ME) MC (\Pythia) and a \NLO ME generator (\Powheg) are used. For the resolved analysis channel, \Pythia 6.425~\cite{pythia} is used with the AUET2B tune~\cite{MC11, MC11c}. For the boosted analysis channel, \Powheg 1.0~\cite{Nason:2004rx, Frixione:2007vw} (patch 4) is used and is interfaced to \Pythia 6.425 for the parton shower, hadronisation, and UE models. Studies of jet substructure and boosted objects have shown that \PowPythia provides a better detector-level description of the internal structure of high-\pT\ jets~\cite{groomedCONF2011}. Comparisons of the boosted topology are also made to the same \Pythia 6.425 MC sample used for the resolved analysis channel in order to evaluate systematic uncertainties. The simulation includes the effect of multiple \pp collisions and is weighted to reproduce the observed distribution of the number of collisions per bunch crossing. Most of the MC samples are processed through a detector simulation~\cite{simulation} based on \geant~\cite{Geant4} and reconstructed in the same manner as data. The only exceptions are the large \Pythia samples that are used for the resolved channel background studies. Due to the very large number of events that are required for these samples, the jets are instead clustered using generator-level particles and their momenta are smeared according to the expected jet energy resolution. With the smearing included, these samples were shown to reproduce the relevant properties of fully-reconstructed data more precisely than is required for this analysis.

\section{Resolved analysis channel}
\label{sec:Resolved}

\subsection{Method and event selection} \label{sec:ResolvedEvtSel}

In the resolved analysis channel, the signal is discriminated from the multijet background by exploiting the large transverse momentum, \pT, of the jets that are produced in gluino decays. The \pT\ of the softest of the leading six jets is used to discriminate the signal from the background. In signal events, the energy is distributed relatively uniformly among each of the six jets. Consequently, the signal is often characterised by six jets each with large \pT, whereas in high-\pT\ QCD multijet background events at least one of the leading six jets is usually produced from soft radiation and is therefore lower in \pT.
Therefore, six jets with an $|\eta| < 2.8$\ are required to pass a certain \pT\ requirement and the observed number of events is compared with expectations. For higher signal masses, the probability of meeting a given jet \pT\ requirement increases due to the increased momentum of the decay products. Thus, it is expected that lower mass signal models will require a lower \pT\ threshold than higher-mass signal models. An optimisation procedure that takes into account both statistical and systematic uncertainties is performed to define the \pT\ requirements which provide the best expected limits in the absence of signal. The \pT\ selection is optimised separately for each generated gluino mass point and three signal regions are chosen. A threshold of $\pT>80$~GeV is chosen for the $\mgluino=100$~GeV gluino mass point, $\pT>120$~GeV for the $\mgluino=200, 300$~GeV gluino mass points, and $\pT>160$~GeV for all higher gluino mass points.

%

Several triggers are used to select events for the signal and control regions studied for this analysis channel. In each case, the triggers are intended to select jets with at least 30~GeV of transverse momentum. This selection has an efficiency greater than 99\% for events in the signal region. A trigger requiring five of these jets was available without prescale during most of the 2011 data taking and is required for events in the signal regions. The interval in which a prescale was active represented less than 1\% of the total data-taking period, and the integrated luminosity that is determined in this paper is corrected to account for lost events. For background estimation from lower jet-multiplicity control regions, triggers requiring only three (four) of these jets are used, with average prescale factors of 530 (57). For all triggers, corrections for the prescales are applied.  

\subsection{Background estimation}
\label{sec:ResolvedBackground}

Standard Model multijet production is the dominant background for the resolved analysis channel. Other backgrounds were considered, including $W+$jets, $Z+$jets, single top, and \ttbar\ production; however they were found to contribute less than 3\% to event yields in all signal and background control regions. The normalisation of the backgrounds is determined starting with the normalisation of data in a control region and using multijet extrapolation factors from simulation to convert to the normalisation in the signal regions. Several different control regions and different extrapolation methods are studied, each giving results consistent with the others. 

%

In the method that is used for the baseline background determination, normalisations are determined starting with the normalisation at lower jet multiplicity in data and then using \Pythia 6 dijet simulation to project into higher jet-multiplicity bins. Such projections are performed using the relation:

\begin{equation}   \label{eq:ProjEq}
 N^{\textnormal{n}-jet}_{data} = N^{\textnormal{m}-jet}_{data} \frac{N^{\textnormal{n}-jet}_{MC}}{N^{\textnormal{m}-jet}_{MC}}
 \end{equation}

\begin{figure*}[t]
\centering
    \includegraphics[height=3.5in]{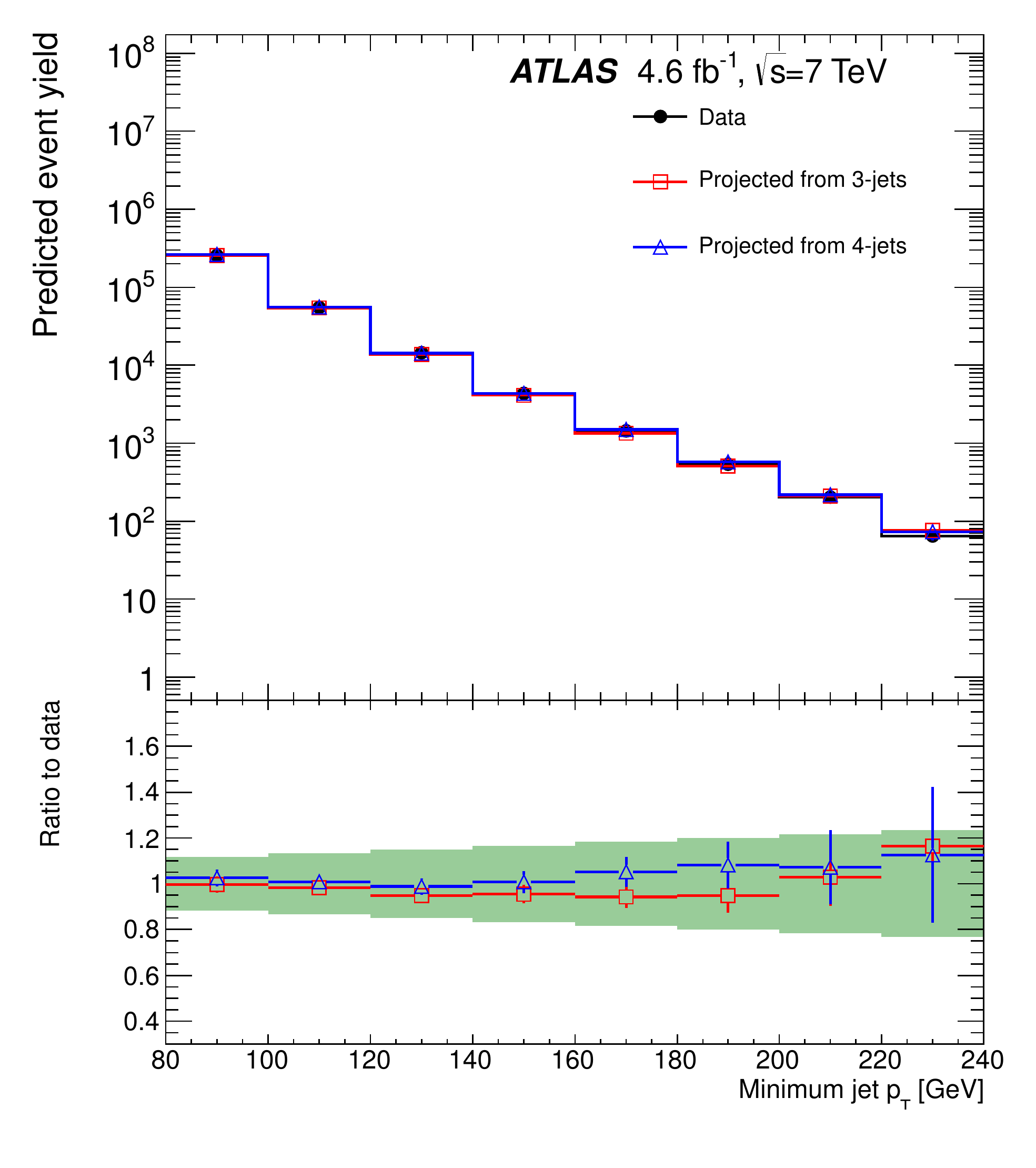}
\caption{Predicted event yield in the 5-jet bin is compared with expectations that are determined by projecting from lower jet multiplicity. The horizontal axis represents the \pT\ selection that is applied when counting jets, and the vertical axis represents the number of events that have exactly five jets with a \pT\ above this threshold. Such comparisons are used to assign a systematic uncertainty to the background normalisation, which is shown as the shaded green band of the ratio plot. The same relative normalisation systematic uncertainty is applied on the background in the signal region.}
\label{fig:5JetPythiaBerendsValidation}
\end{figure*}

Here ``m'' represents the number of jets that are required in a control region, which are then projected to determine the predicted yield when ``n'' jets are required. In the signal region n $\ge 6$ is required. Before performing the final estimation in the signal region, however, the background modelling is first tested by projecting from m $= 3$\ and m $= 4$\ into the signal-depleted n $= 5$ bin. These results are summarised in \figref{5JetPythiaBerendsValidation}. It is seen that the data agree well with background expectations in both cases. In addition to using \equref{ProjEq}, alternative projections are considered in which the simulation is used to project from a lower jet-\pT\ requirement to a higher jet-\pT\ requirement within a given n-jet bin. Both of these projection methods are observed to give consistent background predictions. By examining the largest of the deviations between the data and the predicted background in the n $= 5$\ bin of the data under the jet multiplicity-based extrapolations and the jet \pT-based extrapolation, a systematic uncertainty on the background estimation is chosen. This systematic uncertainty varies between roughly 15\% of the background normalization for loose jet $p_{T}$\ cuts and 25\%  of the background normalization for tight jet $p_{T}$\ cuts.

\begin{figure*}[t]
\centering
    \includegraphics[height=2.7in]{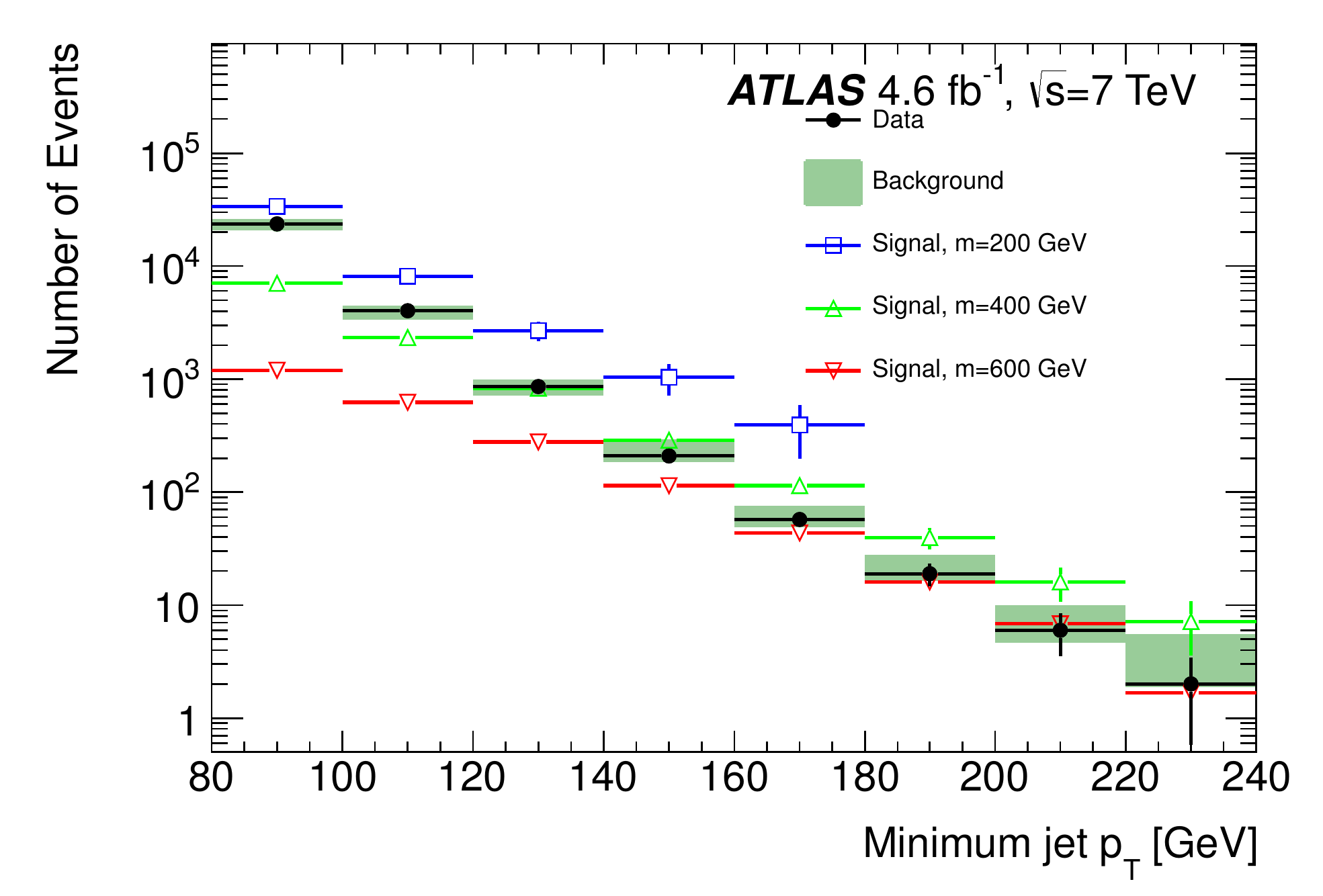}
\caption{Data and the baseline background prediction along with three example signal distributions in the signal region (n $\ge 6$). Background uncertainties include statistical and systematic effects.}
\label{fig:ResolvedBkgDataResultComp}
\end{figure*}

The background in the signal region is determined by using \equref{ProjEq} to project from a 3-jet control region in data into the $\ge 6$-jet bin. This particular projection was chosen because it proved to be the least sensitive to biases from signal contamination in the control region. Projections from the 4-jet bin or from within the $\ge 6$-jet bin from a lower jet \pT\ requirement predict compatible results within uncertainties. The full predicted background and data distributions are overlaid in \figref{ResolvedBkgDataResultComp} along with the predictions from a variety of simulated signal samples. It can be seen that the data agree well with background expectations within uncertainties.

\subsection{Systematic uncertainties}
\label{sec:ResolvedSystematics}

  As discussed in \secref{ResolvedBackground}, the background normalisation systematic uncertainty is chosen to cover the largest discrepancy between data and expectations determined by using the simulation to project into the five-jet bin from control regions of lower jet multiplicity or lower jet transverse momentum. Since this uncertainty is determined directly from comparison to the data, it is considered to cover all systematic uncertainties on the extrapolation factor of \equref{ProjEq}. Cross-checks are run where the simulation is varied within jet energy uncertainties, and variations are found to be well within the uncertainties determined from the data. In addition to the systematic uncertainty on the background there are also statistical uncertainties. These uncertainties come both from data statistical uncertainties in the control region from which the projection begins, and from the statistical uncertainties in the simulation that is used to perform the projection. When projecting into the low jet-\pT\ signal regions that are used in the search for gluino masses below 400 GeV, the systematic uncertainty is much larger than the statistical uncertainty, while when searching for higher signal masses with a tighter jet-\pT\ requirement the statistical uncertainties are larger than the systematic uncertainties. Finally, the systematic uncertainty on potential signal contamination in the background control regions is considered. This systematic uncertainty is evaluated by injecting signal into the data control regions and repeating the background evaluation. The resulting shift in the background is taken as a systematic uncertainty on the background prediction. The results are shown in \tabref{ResolvedNorms}. This uncertainty is only significant for the very low mass gluino models, as these models have both a very large cross-section and predict a significant probability for events to be accepted into the control region.

  The effect of simulation modelling uncertainties on the signal acceptance are also evaluated. The most important sources of uncertainty are due to jet energy modelling. The jet energy resolution (JER) uncertainty has been determined from studies of dijet collisions in the full 2011 dataset~\cite{jespaper2010}. The resulting uncertainties are propagated to this measurement by smearing the jet \pT\ by the appropriate values. Similarly, the uncertainty on the signal acceptance due to jet energy scale (JES) uncertainties is also evaluated by shifting all jet energies coherently. The lower the acceptance of the signal for a given set of selection requirements, the larger the impact of the jet energy scale and resolution uncertainties on the analysis. Depending on the mass point, the JES uncertainty affects the signal acceptance by between 20\% and 30\%, while the effect of the JER uncertainty varies between 5\% and 15\%.

  Systematic uncertainties on the theoretical modelling of the signal properties are also considered. Systematic uncertainties due to theoretical predictions of the inclusive signal cross-section are taken from an envelope of cross-section predictions using different parton distribution function sets and factorisation and renormalisation scales as discussed in~\cite{Kramer:2012bx}. While the inclusive cross-section is determined at NLL+NLO, the probability for collision events to pass selection requirements (``acceptance") cannot be determined in such an accurate manner, so a more conservative systematic estimation is applied. The simulated signal samples use the CTEQ6L1 PDF set~\cite{PDF-CTEQ,cteq6l1}. To determine systematic uncertainties, the signal simulation is reweighted on an event-by-event basis according to the probability for alternative PDFs to produce the generated collision as determined by LHAPDF~\cite{Whalley:2005nh}. It is observed that CTEQ6L1 predicts a lower acceptance for the signal than is predicted by most other PDF sets. A much larger acceptance is predicted by the NNPDF2.0~\cite{Ball:2008by,Ball:2010de} set, while the MSTW2008lo PDF set~\cite{PDF-MRST,Martin:2009iq} predicts an acceptance that is roughly halfway between these two extremes. The MSTW2008lo PDF set has the additional advantage of being determined at LO (which is appropriate for the simulation). The MSTW2008lo PDF set is therefore chosen for the nominal acceptance for the signal samples. Systematic uncertainties are chosen to cover the full difference to the predictions of the acceptance from both the CTEQ6L1 and the NNPDF20 PDF sets, and are added in quadrature to the (smaller) acceptance systematic uncertainties that are determined according to the standard MSTW2008lo prescription. The final signal acceptance uncertainty from PDFs varies between roughly 2\% and 5\%, depending on the signal region.

  Systematic uncertainties on the signal acceptance from QCD radiation are not considered. The reason for this choice is that there is no SM process that contains a colour flow similar to the signal in this analysis due to the presence of colour-epsilon tensors involved in the RPV vertex~\cite{Desai:2011su}. As a consequence, the theoretical understanding of the QCD radiation is less developed than for most other processes, and no clear prescription is available for determining the associated uncertainties. Further, it is important to make these results available in a way that allows them to be applied to six-parton models that may have a different colour flow. The modelling of colour flow and radiation in the signal samples is therefore considered to be part of the model that is analysed in this paper. When reinterpreting the results for other models, it is therefore necessary to account for any differences in colour flow that may arise.


  Other sources of systematic uncertainty are relatively minor. A systematic uncertainty of 3.9\% is included for the integrated luminosity. Since the jet \pT\ requirements are strict, the number of events which fail the trigger requirement but pass all other analysis requirements is less than 1\%. Trigger efficiency systematic uncertainties are therefore also negligible. Similarly, a bias may be present in the background projection factor due to the assumption that all backgrounds are from direct multijet production, when in fact backgrounds such as top-quark and $W+$jets production are also present. As already explained above, these backgrounds are small enough to ignore safely. 

A summary of the expected signal and background events along with the observed data is shown in \tabref{ResolvedNorms}. The systematic uncertainties on the signal are reported separately for each dominant component in \tabref{ResolvedSignalSysts}.

\begin{table}[th]
  \begin{center}
  \begin{tabular}{l|c|c|c|c|c}
\hline  \hline 
Model (\mgluino) & $p_{\textnormal{T,min}}^{6\textnormal{th-jet}}$ & Data & Background & Signal bias [\%] & Signal \\
\hline
$100$~GeV 	&	 80~GeV 	    &    23600 	&	 23500 $\pm$\ 2800 	&	 8.5 	&	 99200 $\pm$\ 20000 \\
$200$~GeV 	&	 120~GeV 	&	 856 	&	 851 $\pm$\ 140 	&	 3.7 	&	 2700 $\pm$\ 500 \\
$300$~GeV 	&	 120~GeV 	&	 856 	&	 851 $\pm$\ 140 	&	 1.0 	&	 1460  $\pm$\ 240\\
$400$~GeV 	&	 160~GeV 	&	 57 	    &	 62 $\pm$\ 13 	&	 0.8 	&	 110 $\pm$\ 13\\
$500$~GeV 	&	 160~GeV 	&	 57 	    &	 62 $\pm$\ 13 	&	 0.3 	&	 67 $\pm$\ 9\\
$600$~GeV 	&	 160~GeV 	&	 57 	    &	 62 $\pm$\ 13 	&	 0.1 	&	 43 $\pm$\ 7\\
$800$~GeV 	&	 160~GeV 	&	 57 	    &	 62 $\pm$\ 13 	&	 0.0 	&	 20 $\pm$\ 3\\
\hline \hline 
  \end{tabular}
  \caption{Number of events expected for the background and signal for each of the models in the resolved gluino search along with the number of observed events. Most of the uncertainties on the background and signal models are included in columns four and six. The one exception is the bias of the background normalization that results from signal contamination in the background control regions. The fractional bias resulting from this effect is shown in the fifth column.} \label{tab:ResolvedNorms}
  \end{center}
\end{table}
\begin{table}[th]
  \begin{center}
  \begin{tabular}{l|c|c|c|c|c}
\hline \hline 
 Source 	& $\mgluino = 100$~GeV & 200~GeV & 400~GeV & 600~GeV & 800~GeV \\
\hline
Jet Energy Scale 	&	20	&	16	&	11	&	18	&	13\\
Jet Energy Resolution 	&	2.7	&	12	&	3.5	&	2.8	&	1.5 \\
PDFs  & 4.9 & 4.1 & 2.6 & 4.7 & 4.7\\
Total 	&	21	&	20	&	12	&	19	&	14\\
\hline  \hline 
  \end{tabular}
  \caption{Largest relative systematic uncertainties (in \%) on the signal acceptance for the resolved analysis at each gluino mass point. Please note that the values of these uncertainties do not evolve in a fully continuous way because the selection cuts are tighter for the higher mass points. In general, tightening the selection cuts raises these uncertainties while going to a higher mass value for a given selection cut lowers them.} \label{tab:ResolvedSignalSysts}
  \end{center}
\end{table}

\section{Boosted analysis channel}
\label{sec:Boosted}
\subsection{Method and event selection}

A complementary method is adopted for the search in the low gluino mass region wherein gluinos may be produced with a large boost ($\pT\gtrsim2\times \mgluino$). In such a topology, the three quarks from each gluino decay can be very collimated and therefore reconstructed as a single large-radius jet with a distance parameter of $R=1.0$. The advantage of this method is that the single-jet invariant mass and properties of the internal structure of such a jet provide discriminants against the large SM multijet background. The signal region definition is approximately orthogonal to that of the resolved channel described above and carries nearly independent experimental systematic uncertainties. As a result, the boosted technique provides not only a well-motivated cross-check for a challenging all-hadronic search, but it also establishes the use of jet substructure with boosted objects for future SUSY searches.


Events are selected using either a high \pT\ single jet trigger ($\pT>240$~GeV) or a slightly lower \pT\ single jet trigger ($\pT>100$~GeV) with an additional requirement on the total summed \pT\ of jets reconstructed in the trigger system. The offline jet selection ($\pT>350$~GeV or $\pT>200$~GeV, for the two trigger options) is based on \antikt\ jets with a radius $R=1.0$ in order to maximise the efficiency for moderately boosted massive gluinos. For the complete offline selection criteria, including the requirement on the jet multiplicity described below (see \tabref{signal-control:cuts}), the inefficiency of the trigger for the boosted gluino signal is less than 1\%. Jets are found using the same locally calibrated \topos{} described in \secref{ResolvedEvtSel}. The energy of the resulting \largeR jets is calibrated with a MC-derived calibration factor~\cite{groomedCONF2011} that is dependent on the uncalibrated jet \pT\ and \eta. In addition to the energy calibration, a mass calibration is applied that accounts for differences between the particle- and reconstructed-level jet invariant mass observed in MC simulation. The energy and mass scale uncertainties of the calibrated jets are determined using \insitu\ measurements of inclusive jet samples and are found to be approximately 4\% and 5\%, respectively.

\begin{figure}[!ht]
    \centering
    \subfigure[Signal region 1 preselection]{
      \includegraphics[width=0.47\textwidth]{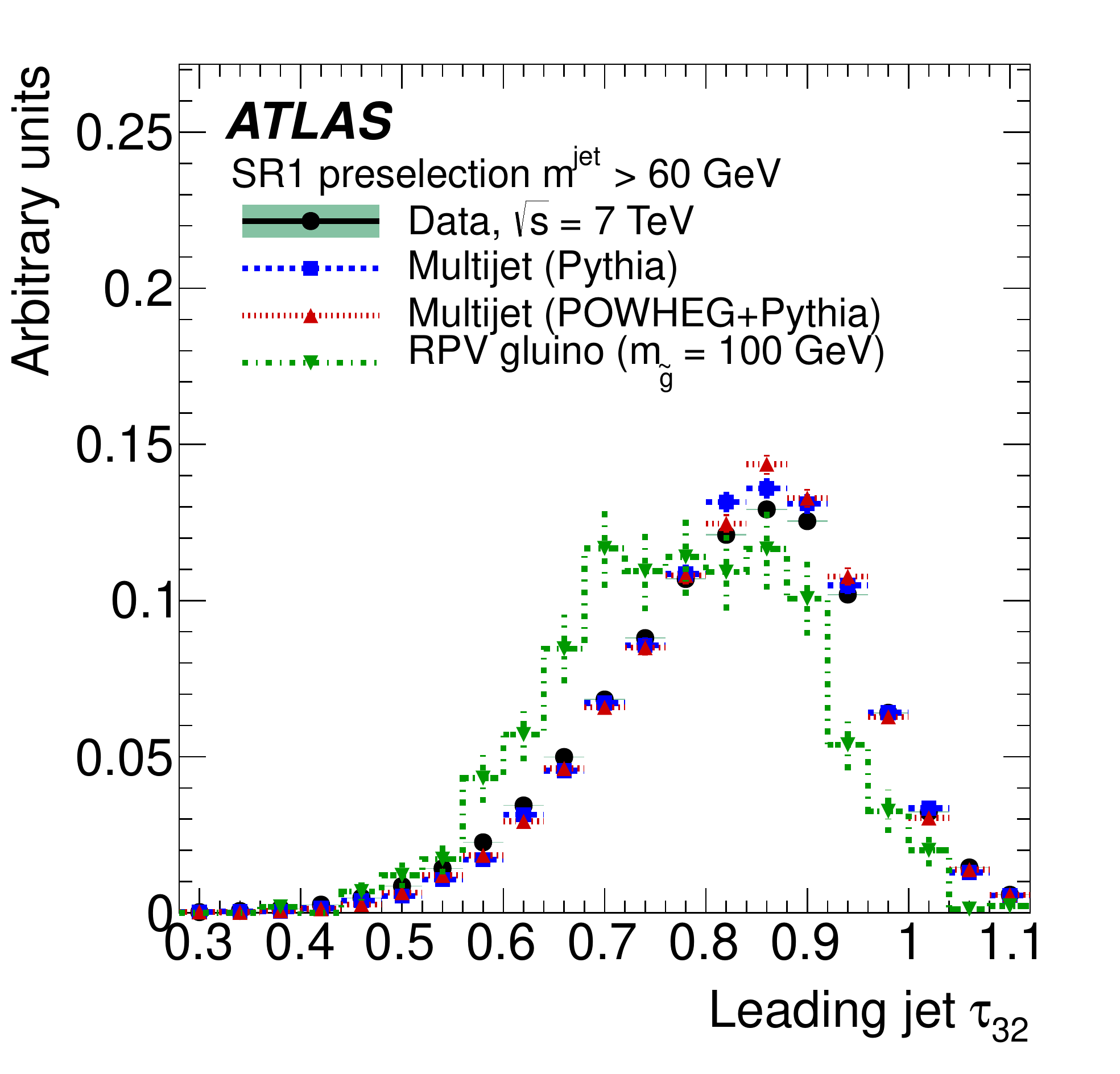}
      \label{fig:results:SR:tau}
    }
    \subfigure[Signal region 1]{
      \includegraphics[width=0.47\textwidth]{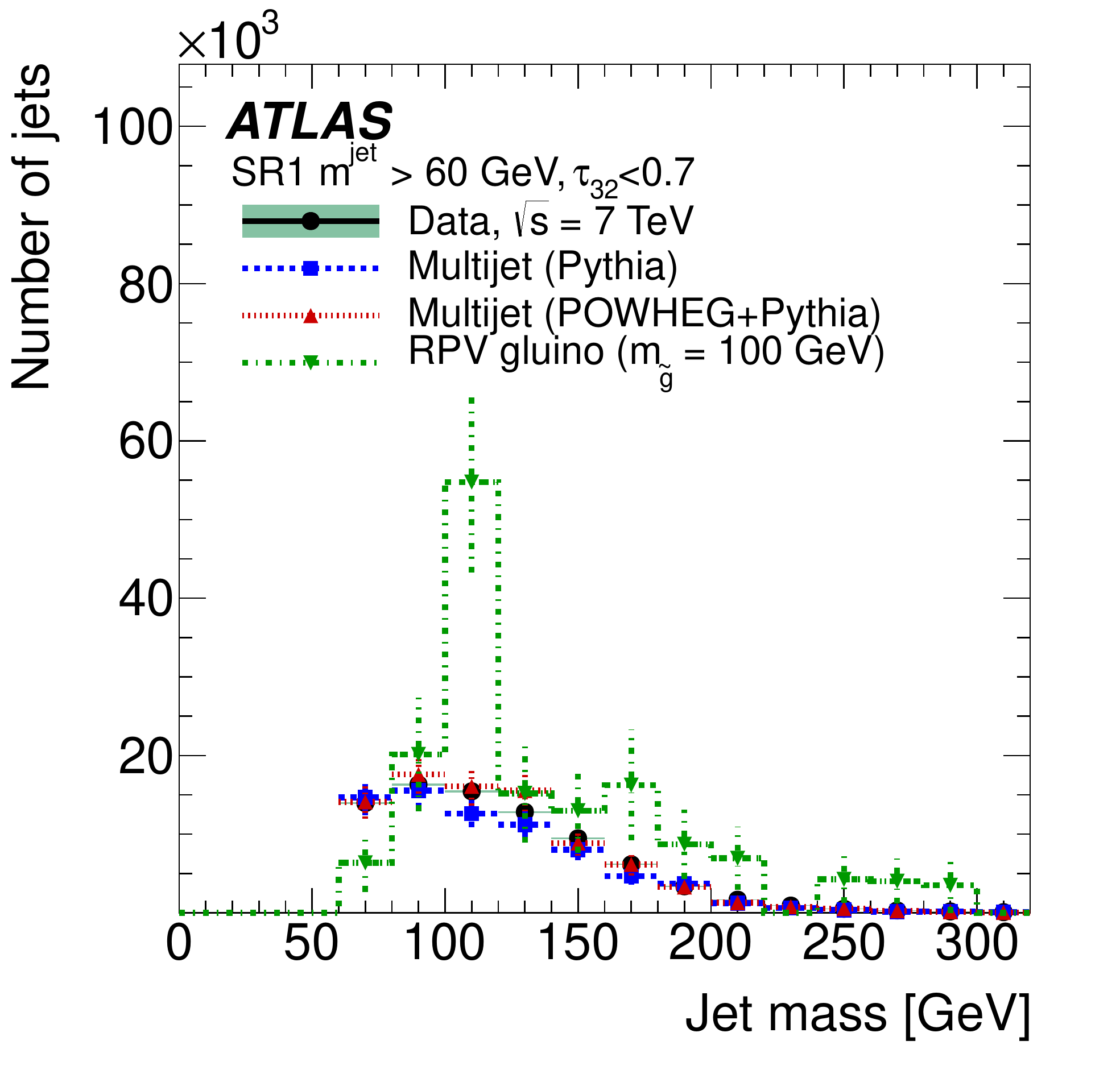}
      \label{fig:results:SR:mass}
    }
    \subfigure[Signal region 2 preselection]{
      \includegraphics[width=0.47\textwidth]{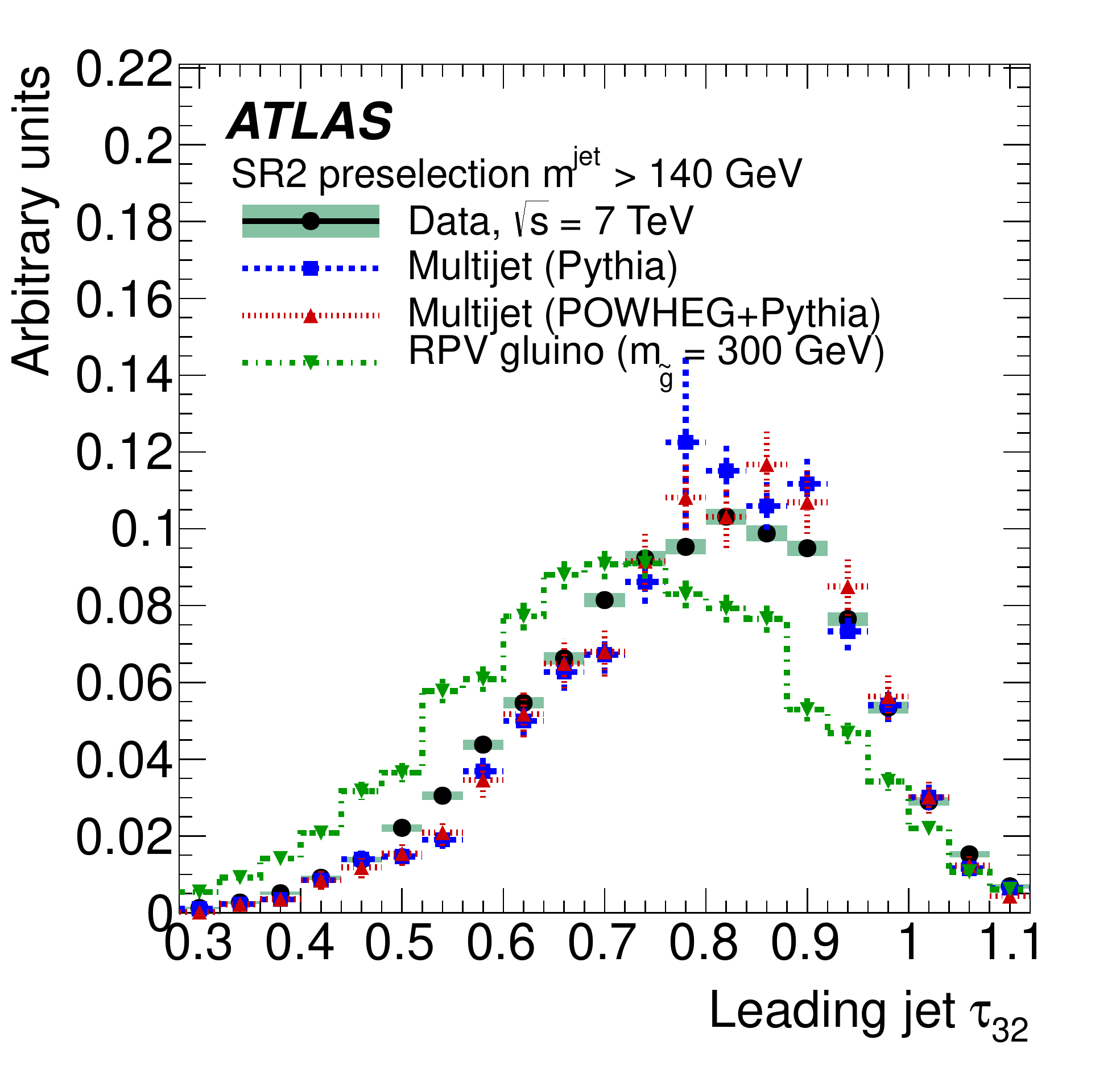}
      \label{fig:results:SR2:tau}
    }
    \subfigure[Signal region 2]{
      \includegraphics[width=0.47\textwidth]{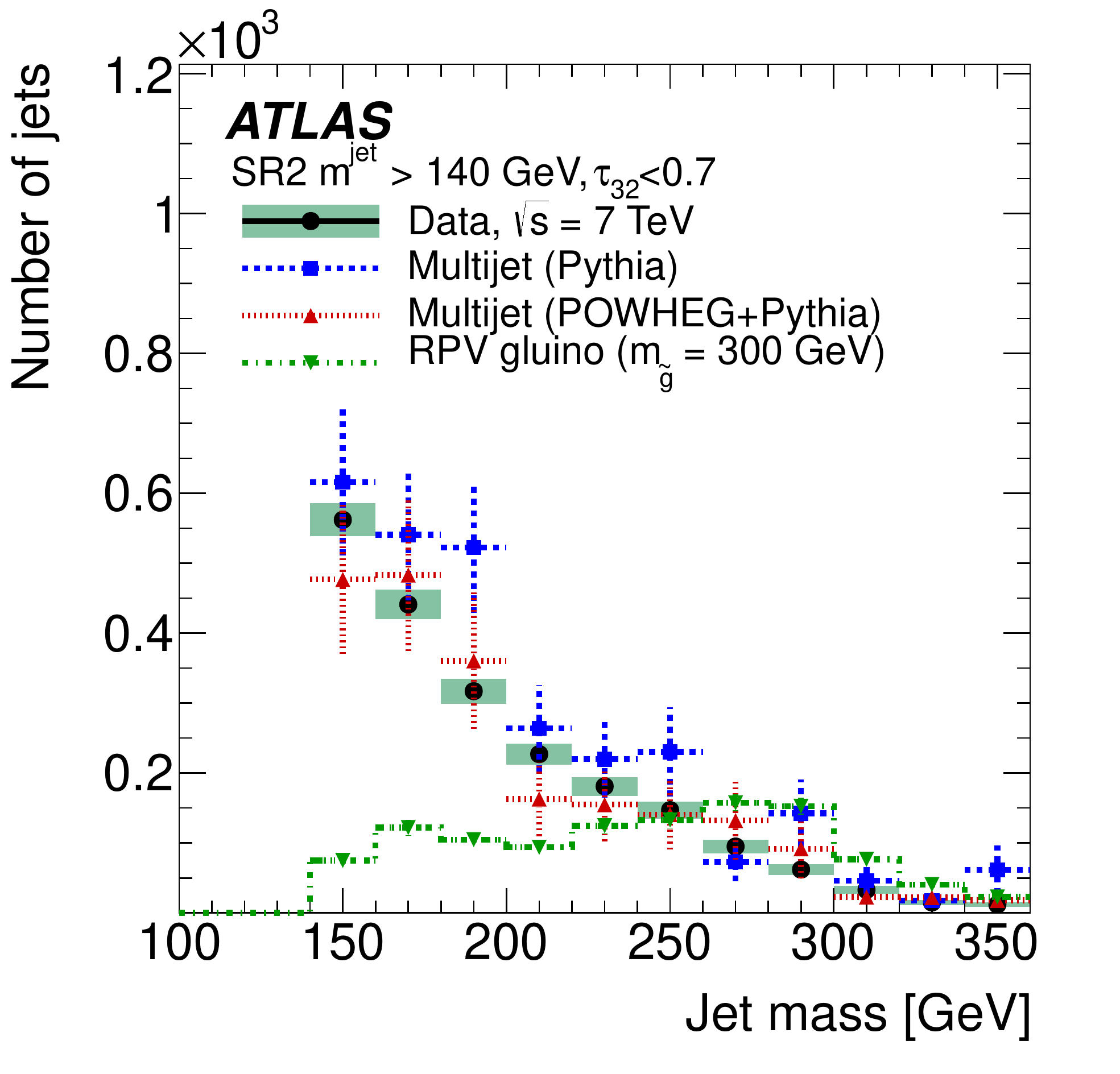}
      \label{fig:results:SR2:mass}
    }

  \caption{In the lower mass signal region (SR1), the distributions of \subref{fig:results:SR:tau} jet \tauThrTwo for the two leading jets in each event with $\mjet>60$~GeV and \subref{fig:results:SR:mass} jet mass (\mjone and \mjtwo) for jets with $\tauThrTwo<0.7$ are shown for the data, the signal $\mgluino=100$~GeV, and the background MCs for comparison. In the higher mass signal region (SR2), the same distributions of \subref{fig:results:SR2:tau} \tauThrTwo and \subref{fig:results:SR2:mass} jet mass are shown, but in this case for $\mgluino=300$~GeV. In each case, the data are compared to the two MC models used to estimate the correlation correction factor, $\alpha$, for the background extrapolation.}
  
  \label{fig:boostedresults:SR}
  \end{figure}

In order to provide discrimination against multijet events containing jets with a large mass, we use a jet shape variable that is sensitive to the $N$-body structure expected from a jet containing the three decay products of a light gluino. The ``$N$-subjettiness" variables \tauN~\cite{Thaler:2010tr, Thaler:2011gf} provide this sensitivity as they relate to the subjet multiplicity on a jet-by-jet basis. The $\tau_N$ variables are calculated by re-clustering all of the \topo\ constituents of the jet with the exclusive \kt algorithm~\cite{Catani1993} and requiring $N$ subjets to be found. These $N$ subjets define axes within the jet, around which the jet constituents may be concentrated. 
  The variables $\tau_N$ are defined in \equref{grooming_nsubj} as the sum over all constituents ($k$) of the jet:

    \begin{equation}
      \tauN = \frac{1}{d_{0}} \sum_k p_{\mathrm{T}k} \times \text{min}(\delta R_{1k}, \delta R_{2k},...,\delta R_{Nk})~\text{,~~with} 
       ~~~d_{0}\equiv\sum_{k} p_{\mathrm{T}k}\times R
      \label{eq:grooming_nsubj}
    \end{equation}
%
  where $R$ is the jet radius parameter in the jet algorithm, $p_{\mathrm{T}k}$ is the \pT\ of constituent $k$ and $\delta R_{ik}$ is the distance from the subjet $i$ to constituent $k$.
  Using this definition, $\tau_N$ characterises how well a jet can be described as containing $N$ or fewer \kt\ subjets. Constituents localised near the axes of the subjets will result in a relatively smaller value of $\tau_N$, thereby categorizing such a jet as likely to be comprised of at most $N$ subjets. 
  The ratio $\tau_3/\tau_2$, written also as \tauThrTwo, is used to provide discrimination between jets formed from the parton shower of light quarks or gluons and jets containing three hadronic decay products from boosted gluinos. 
  A value $\tauThrTwo \simeq 1$ corresponds to a jet that is very well described by two subjets and $\tauThrTwo \simeq 0$ implies a jet that is much better described by three subjets than one or two. The distribution of \tauThrTwo for signal and background MC events, as well as that observed in the data, is shown in \figref{boostedresults:SR} in \secref{signal-control:unc}.

Following the jet reconstruction, and after the calculation of \tauThrTwo, the trimming algorithm~\cite{Krohn2010} is used to remove soft energy depositions from the jet that can degrade the jet properties in the presence of \pileup or significant underlying event contamination. The procedure uses the inclusive \kt algorithm~\cite{Ellis1993} to create subjets of size $\drsub=0.3$ from the constituents of a jet. Any subjets with $\pti/\ptjet < \fcut$ are removed, where \pti\ is the transverse momentum of the $i^{th}$ subjet, and $\fcut=0.05$ is determined to be an optimal setting for improving the mass resolution in the presence of \pileup~\cite{groomedCONF2011, groomedJetPileup2011}. The remaining constituents form the trimmed jet. The invariant mass of these \largeR, trimmed jets is then calculated from the energies and momenta of the constituents contained within the jet after the trimming procedure.


Events containing pair produced boosted gluinos that decay into three collimated quarks are characterised by the presence of two massive \largeR jets that each contain substructure representative of a massive three-body decay. The substructure ``tag'' is defined by $\tauThrTwo<0.7$ which has been determined by optimising the selection based on the signal-to-background ratio expected from MC simulation studies. The efficacy and MC modelling of this approach has been validated using events containing high-\pT\ pair produced top quarks with one top quark decaying leptonically and the second decaying hadronically~\cite{groomedCONF2011}. The invariant mass of single \largeR jets containing the fully hadronic three-body decay of a top quark and the \tauThrTwo distribution are both  well described by the MC simulation in terms of shape and rate. 

In addition to the jet-level mass and substructure-based signal discrimination, the event-level jet multiplicity using small-radius $R=0.4$ jets (\NjetStd) with $\ptjet>30$~GeV also provides discrimination power. Events containing highly boosted gluinos are nonetheless expected to contain at least four individual small-radius jets due to partial separation of the decay products and hard, final-state radiation (FSR). Consequently, both jet-level and event-level observables are available for signal and background discrimination. The multijet background exhibits a maximum at $\NjetStd=3$, as expected from high-\pT\ dijet events, whereas the signal peaks near $\NjetStd=4-5$ due to the multiple hard partons in the final state including FSR. The event-level selection $\NjetStd\geq4$ is chosen as a result of this observation. 

\begin{table}[!ht]
  \begin{center}
  \begin{tabular}{l | c | c | c}
    \hline \hline
     Selection                     & Baseline Selection  & SR1 & SR2       \\ \hline
     Small-$R$ ($R=0.4$) jet \ptjet & $\ptjet>30$~GeV & $\ptjet>30$~GeV & $\ptjet>30$~GeV      \\ 
     \LargeR ($R=1.0$) jet \ptjet & $\ptjet>200$~GeV & $\ptjet>200$~GeV & $\ptjet>350$~GeV  \\
     Scalar sum $\sum_{i=1}^{\NjetStd=4}\ptjet$  & (---) & 600~GeV & (---)    \\ \hline
     Small-$R$ jet multiplicity & (---) & $\NjetStd\geq4$ & $\NjetStd\geq4$        \\ 
     \LargeR jet multiplicity  & $\Njet\geq2$ & $\Njet\geq2$ &  $\Njet\geq2$      \\ \hline
     \LargeR jet mass & (---) & $\massjet_{\jone,\jtwo}>60$~GeV & $\massjet_{\jone,\jtwo}>140$~GeV  \\
     \LargeR jet \tauThrTwo   & (---) & $\tauThrTwo<0.7$ & $\tauThrTwo<0.7$      \\
    \hline \hline
  \end{tabular}
  \end{center}
  
  \caption{Baseline and signal selection criteria at both the event-level and jet-level for signal region one (SR1) and two (SR2).}
  
  \label{tab:signal-control:cuts}
\end{table}

\Tabref{signal-control:cuts} presents the baseline event and object selections, as well as the additional selection criteria that define the signal regions studied for the analysis. The signal region (SR) optimised for lower mass (SR1) requires a lower jet \pT\ threshold and includes an additional requirement on the total scalar sum of jet momenta using the four leading small-radius jets ($\sum_{i=1}^{\NjetStd=4}p_{{\rm T},i}^{\rm jet}$). In \tabref{signal-control:cuts}, the ``leading'' jet refers to both the first (\jone) and the second (\jtwo) \largeR jets in the event, ordered according to \ptjet. The higher mass signal region (SR2) only requires a high-\pT\ leading \largeR jet in the event with $\ptjet>350$~GeV, and is optimised for signal models with $\mgluino>200$~GeV.

\subsection{Background estimation}

Standard Model multijet production is the dominant background in this approach. The backgrounds in the signal regions described above are estimated using an ``ABCD method" wherein event yields in orthogonal control regions are used to predict the total number of events expected in the signal region. The control region definitions rely on the inversion of the signal region selection criteria which are defined in \tabref{signal-control:cuts}. In particular, inclusive events and events with low \NjetStd are used to assess the description of the mass and substructure observables for high-mass \largeR jets.

Three primary control regions (CR-$A, B, C$) are used to estimate the background. The selections applied are summarised in \tabref{signal-control:ABCD}. Control region $A$ (CR-$A$) is comprised of low-mass jets ($\massjet<60$~GeV or $\massjet<140$~GeV, for SR1 and SR2, respectively) with no substructure criteria applied, CR-$B$ contains a single high-mass leading jet ($\massjet>60$~GeV or $\massjet>140$~GeV), and a low-mass subleading jet ($\massjet<60$~GeV or $\massjet<140$~GeV). In addition, the leading jet in CR-$B$ has a substructure tag ($\tauThrTwo<0.7$), and CR-$C$ is defined in a very similar way to CR-$B$, but where the subleading jet is massive and contains a substructure tag, whereas the leading jet is required to have a low mass. 

\begin{table}[!ht]
  \begin{center}
  \begin{tabular}{l | l l p{4cm}}
    \hline \hline
     Region     & Jet (\jone) selections & Jet (\jtwo) selections & Description \\\hline\hline
          
     \multirow{2}{*}{CR-$A$}     & \multirow{2}{*}{$\massjet<\mThresh$}                  & \multirow{2}{*}{$\massjet<\mThresh$}  & Low-mass jets, \,\,\,\,\,\,\,\,\,\,\,\,\,\,\,\,\,\,\,\,\, to validate \tauThrTwo shape  \\\hline 
     
     \multirow{2}{*}{CR-$B$} & $\massjet>\mThresh$                  & \multirow{2}{*}{$\massjet<\mThresh$}   & Signal-like leading jet,  \\   
                        & $\tauThrTwo<0.7$                    &       & to validate \massjet                      \\\hline
     
     \multirow{2}{*}{CR-$C$} & \multirow{2}{*}{$\massjet<\mThresh$} & $\massjet>\mThresh$  & Signal-like subleading    \\   
                        &                                     & $\tauThrTwo<0.7$     & jet, to validate \massjet       \\ 
    \hline \hline
  \end{tabular}
  \end{center}
  
  \caption{Definition and description of the four primary control regions used to estimate the backgrounds using the ABCD method. $\mThresh = 60$ (140) GeV for SR1 (SR2).}
  
  \label{tab:signal-control:ABCD}
\end{table}

The background estimation is designed to be performed directly from the data with minimal input from multijet MC simulation. First, the normalisation for the leading \largeR jet mass distribution is obtained using orthogonal control regions by computing the ratio of the number of events in CR-$B$ to CR-$A$, multiplied by the number of events in CR-$C$, as given in \equref{abcd}. Second, a correlation correction factor, $\alpha$, defined in \equref{alpha}, is necessary to properly handle correlations between the signal region and control region estimates. This correlation correction factor is evaluated from \PowPythia MC samples in order to avoid potential signal contamination.

\begin{equation}
  N_{SR} = N_{{\rm CR-}C}\times\left(\frac{N_{{\rm CR-}B}}{N_{{\rm CR-}A}}\right)\times\alpha
  \label{eq:abcd}
\end{equation}

\begin{equation}
  \alpha = \left.\left(\frac{N_{SR}\, /\, N_{{\rm CR-}C}}{N_{{\rm CR-}B}\, /\, N_{\rm CR-A}}\right)\right|_{\rm MC}
  \label{eq:alpha}
\end{equation}

\begin{figure}[tp]
  \centering
  \subfigure[Data]{
    \includegraphics[width=0.45\textwidth]{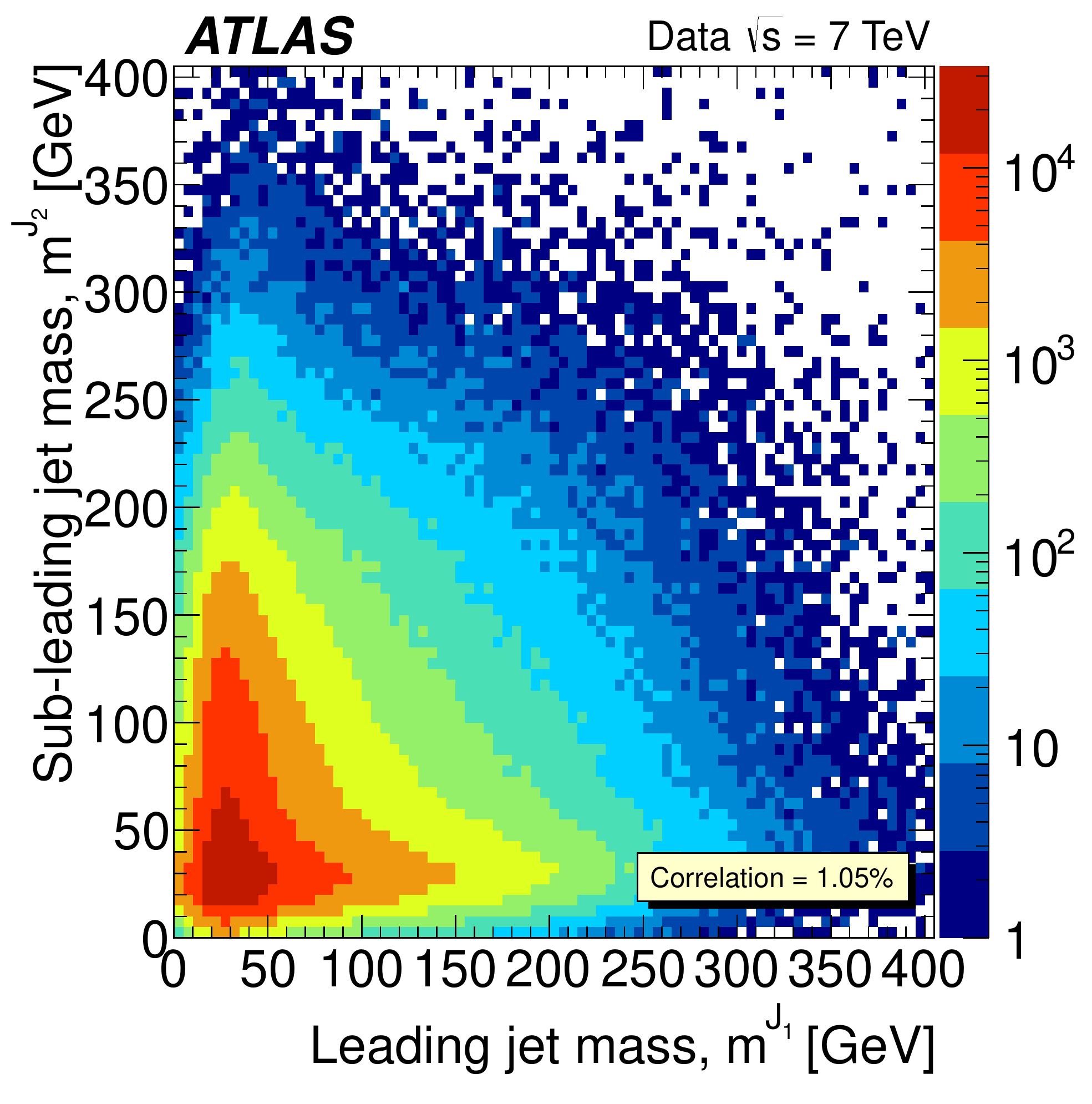}
    \label{fig:signal-control:masscorr:data}
  }
  \subfigure[\PowPythia]{ 
    \includegraphics[width=0.45\textwidth]{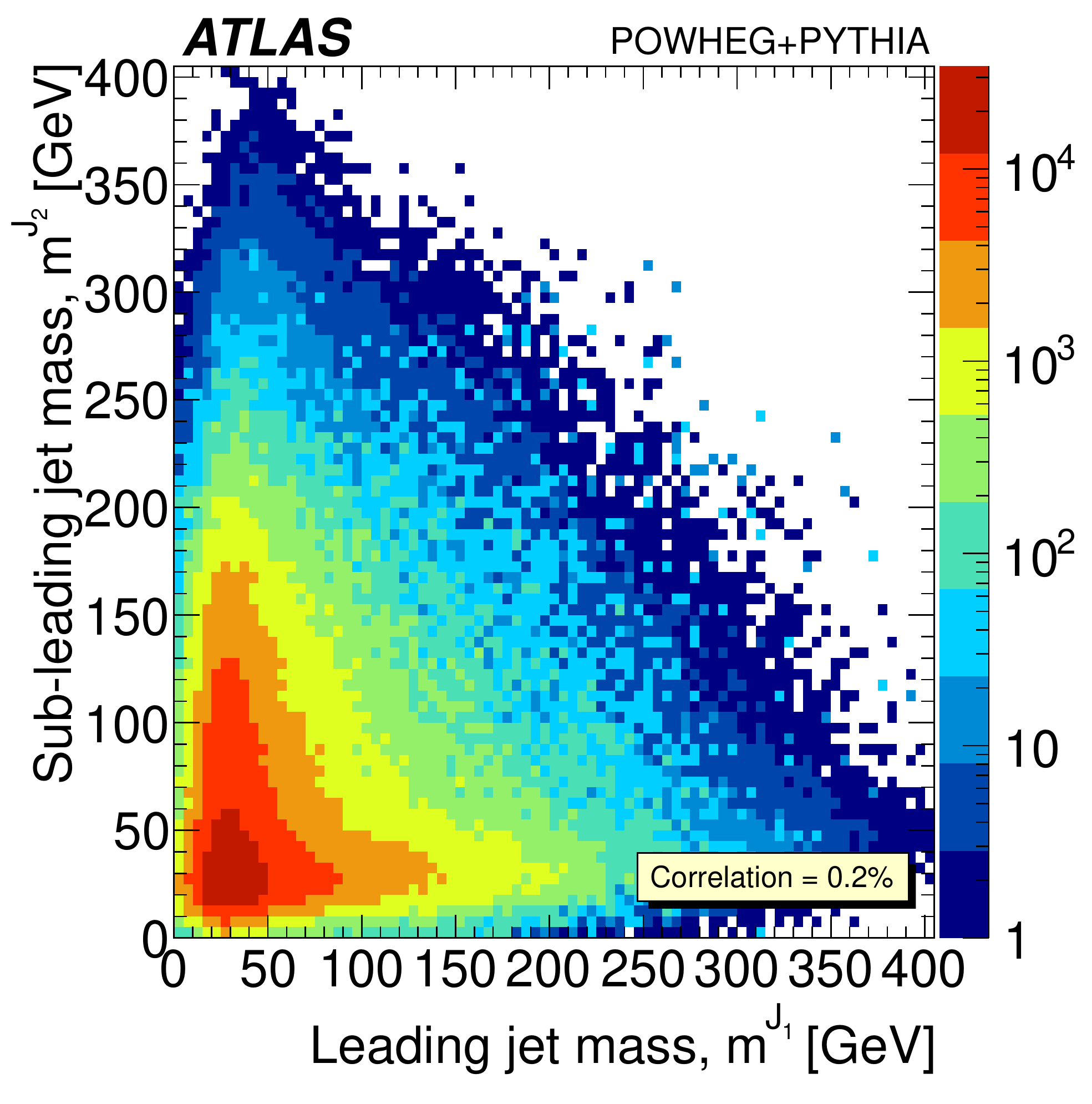}
    \label{fig:signal-control:masscorr:powheg}
  } \\
  \subfigure[Data, $\massjet>100$~GeV]{
    \includegraphics[width=0.45\textwidth]{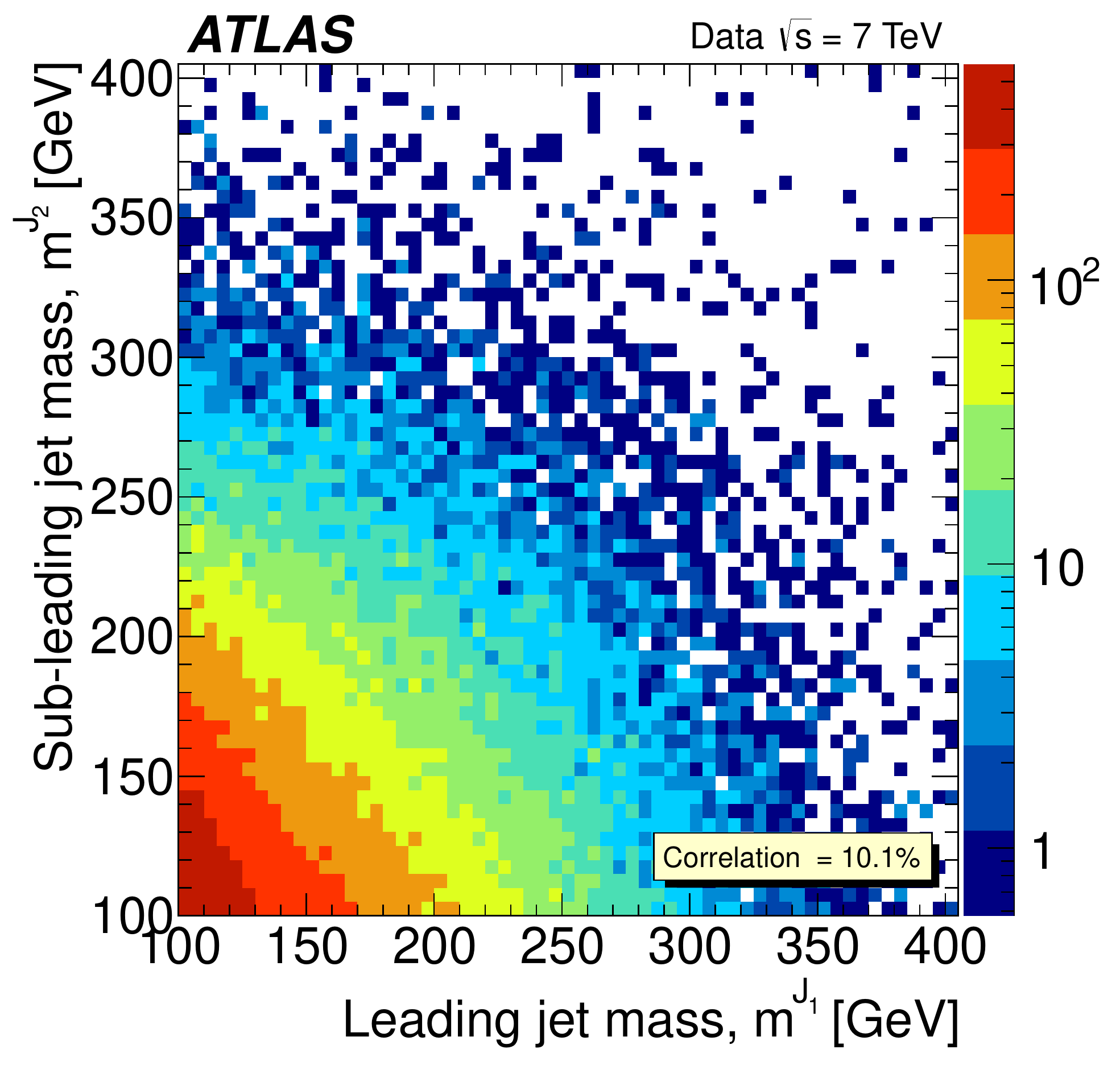}
    \label{fig:signal-control:masscorr100:data}
  }
  \subfigure[\PowPythia, $\massjet>100$~GeV]{ 
    \includegraphics[width=0.45\textwidth]{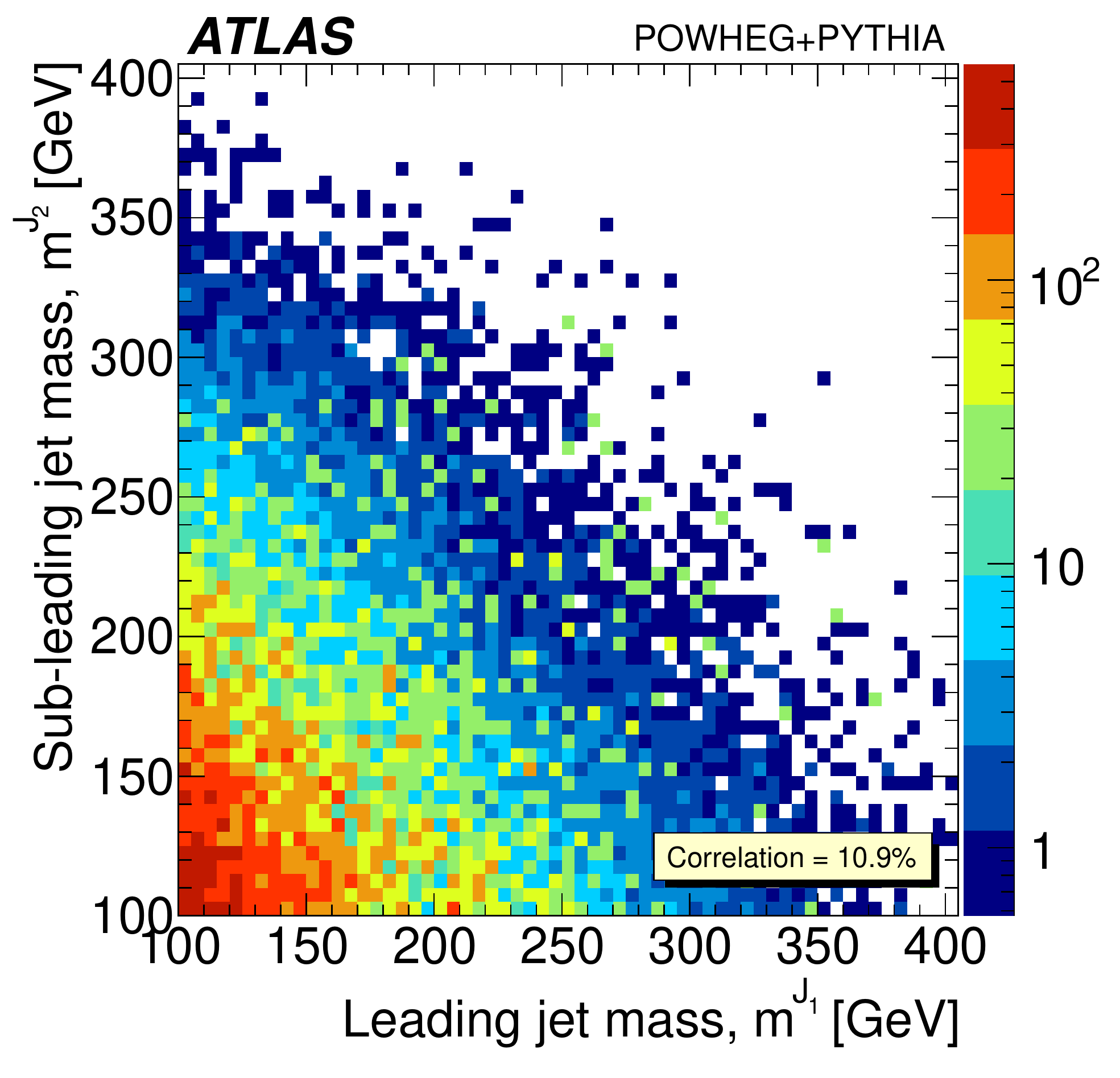}
    \label{fig:signal-control:masscorr100:powheg}
  }
      
    \caption{Distributions of the first leading and subleading (in \ptjet) jet masses from which the correlation coefficients ($\rho$) are determined in \subref{fig:signal-control:masscorr:data} data ($\rho=1.05\%$), \subref{fig:signal-control:masscorr:powheg} \PowPythia MC samples ($\rho=0.2\%$), \subref{fig:signal-control:masscorr100:data} data with $\massjet>100$~GeV ($\rho=10.1\%$), and \subref{fig:signal-control:masscorr100:powheg} \PowPythia MC samples with $\massjet>100$~GeV ($\rho=10.9\%$).
    }
  
  \label{fig:signal-control:masscorr}   
      
\end{figure}

This effect and modelling of the leading jet-mass correlations are studied using the baseline event selection (i.e. no selection on \NjetStd, or \tauThrTwo) via the correlation coefficient\footnote{The correlation coefficient, $\rho$, is calculated from the covariance of the two observables, \mjone and \mjtwo, and the root-mean-square ($RMS$) of each observable using the expression $\rho={\rm cov}(\mjone,\mjtwo)/\left[RMS(\mjone)\times RMS(\mjtwo)\right]$. A value $\rho=0$ indicates no correlation.}, $\rho$. A slightly larger correlation between the two leading jet masses is present in data (1.05\%) than predicted by the \PowPythia MC samples (0.2\%), as shown in \figsref{signal-control:masscorr:data}{signal-control:masscorr:powheg}. However, when restricting the mass range to $\massjet>100$~GeV, as in \figsref{signal-control:masscorr100:data}{signal-control:masscorr100:powheg}, the correlation coefficient is observed to be 10.1\% (10.9\%) in data (MC). Given this relatively good agreement, a prediction for $\alpha$ is made using the \PowPythia MC samples.

The expected background in the signal regions as determined from the ABCD method described above, as well as the observed event yields and the predicted signal yield for the two low-mass gluino models, are summarised in \tabref{signal-control:yields}. The systematic uncertainties on the background prediction and the expected signal yield described in \secref{signal-control:unc} are included in \tabref{signal-control:yields}.

\begin{table}[!ht]
  \begin{center}
  \begin{tabular}{l|c|c|c|c|c}
  \hline \hline 

Model (\mgluino) & \mThresh & Data  & Background     & Signal Bias [\%] & Signal \\ \hline
$100$~GeV        & 60~GeV   & 40683 & $42400\pm9700$ & 65               & $77900 \pm 16000$ \\
$200$~GeV        & 140~GeV  & 1059  & $860  \pm460$  & 31               & $2400  \pm 670$   \\
$300$~GeV        & 140~GeV  & 1059  & $860  \pm460$  & 9                & $590   \pm 55$    \\

  \hline \hline 
  \end{tabular}
  \caption{Number of events expected for the background and signal for each of the models in the boosted gluino search along with the amount of observed data. The uncertainties on the background prediction and the expected signal yield are included. The bias of the background normalization that results from signal contamination in the background control regions is shown separately in the fifth column.} 
  \label{tab:signal-control:yields}
  \end{center}
\end{table}

\subsection{Systematic uncertainties}
\label{sec:signal-control:unc}

The primary systematic uncertainties affecting this analysis are those related to the kinematic scales of the jets used to define the signal regions (mass and \pt) as well as those that affect the background estimation method. The systematic uncertainties on the measurement of the jet mass and \pT\ are evaluated using inclusive jet measurements, as well as samples enriched in boosted $W$ bosons and top quarks~\cite{groomedCONF2011}. For the \largeR jets used in this analysis, the typical jet mass scale uncertainties are approximately 5\%, whereas the energy scale uncertainties are approximately 4\%. These impact the jet mass distribution and the correlation correction factor, $\alpha$, used to extrapolate the background estimates from the control region into the signal region. 

The difference between $\alpha$ evaluated using \PowPythia as compared to \Pythia is symmetrised and taken as a systematic uncertainty. Furthermore, additional systematic uncertainties on the determination of $\alpha$ itself are evaluated using the \PowPythia MC samples by varying the jet energy and mass scales. These variations are performed in each signal region and control region separately. The energy scale is considered as uncorrelated with the mass scale, whereas the two are each considered as correlated between the leading and subleading jets in the event. The impact on the determination of $\alpha$ is:

\begin{eqnarray}
 \alpha_{\mjet>60\GeV} &=& 0.54 \pm 0.05\; {\rm (stat.)} \pm 0.03\; {\rm (syst.)} \pm 0.08\; {\rm (MC\; syst.)} \nonumber \\
 \alpha_{\mjet>140\GeV} &=& 0.27 \pm 0.04\; {\rm (stat.)} \pm 0.03\; {\rm (syst.)} \pm 0.08\; {\rm (MC\; syst.)} \nonumber
\end{eqnarray}

\noindent where the systematic uncertainty due to the jet energy and mass scales is separated from the systematic uncertainty due to the MC comparisons between \Pythia and \PowPythia. The impact of contamination due to pair produced top quarks contaminating the signal or control regions has been explicitly evaluated and is observed to be less than 5\% (10\%) for the low (high) gluino mass signal region, which is to be compared to an overall systematic uncertainty of 23\% (53\%).

The impact of the kinematic scale variations and effect of PDF set variations on the signal acceptance are assessed as systematic uncertainties on the signal yield for each gluino mass hypothesis. The systematic uncertainty on the signal acceptance due to the PDF set variation is evaluated independently for the boosted topology selection in the same manner as described in \secref{ResolvedSystematics} since there is the potential that the two selections are affected differently. \Tabref{signal-control:signalsyst} summarises the systematic uncertainties on the signal yield that are included in the final results. 

\begin{table}[!ht]
  \begin{center}
  \begin{tabular}{ l | c | c | c  }
    \hline \hline
Source              & $\mgluino=100$~GeV    & $\mgluino=200$~GeV & $\mgluino=300$~GeV\\ \hline
Jet energy scale (JES) & $+8.7/-6.4$           & $+10/-8.9$         & $+5.8/-5.5$ \\
Jet mass scale (JMS)   & $\lesssim$1           & $+15/-4.2$         & $+4.7/-4.7$ \\
Total JES+JMS          & $+8.7/-6.4$           & $+18/-9.8$         & $+7.5/-7.2$ \\
PDFs                   & $+5.1/-2.1$           & $+2.3/-3.0$        & $+4.0/-4.0$ \\
MC statistics          & $18$                  & $22$               & $4.1$       \\ \hline
Total                  & $+21/-19$             & $+28/-24$          & $+9.4/-9.2$   \\      
    \hline \hline
  \end{tabular}
  \end{center}
  
  \caption{Largest relative systematic uncertainties (in \%) on the signal acceptance for the boosted analysis.
  }
  
  \label{tab:signal-control:signalsyst}
\end{table}

The \tauThrTwo distribution for the two leading jets in each event with $\mjet>60$~GeV and $\mjet>140$~GeV, as well as the  mass distribution for leading and subleading jets (\mjone and \mjtwo) with $\tauThrTwo<0.7$, are shown in \figref{boostedresults:SR}. In each case the use of the \tauThrTwo observable improves the signal-to-background ratio by approximately a factor of 3.5. This improvement is not quite as large as that expected from studies of boosted top quarks~\cite{Krohn2010, BOOST2011} due to the relatively soft requirement on the gluino \pT\ and the different colour structure of the final state. After the \tauThrTwo selection, the trimmed jet mass distributions for the two gluino mass hypotheses shown in \figref{boostedresults:SR} provide considerable discrimination from the SM QCD multijet background, which is characterised by a smoothly falling distribution.

\section{Results}
\label{sec:Results}
%

\begin{figure*}
  \centering
      \includegraphics[width=0.9\textwidth]{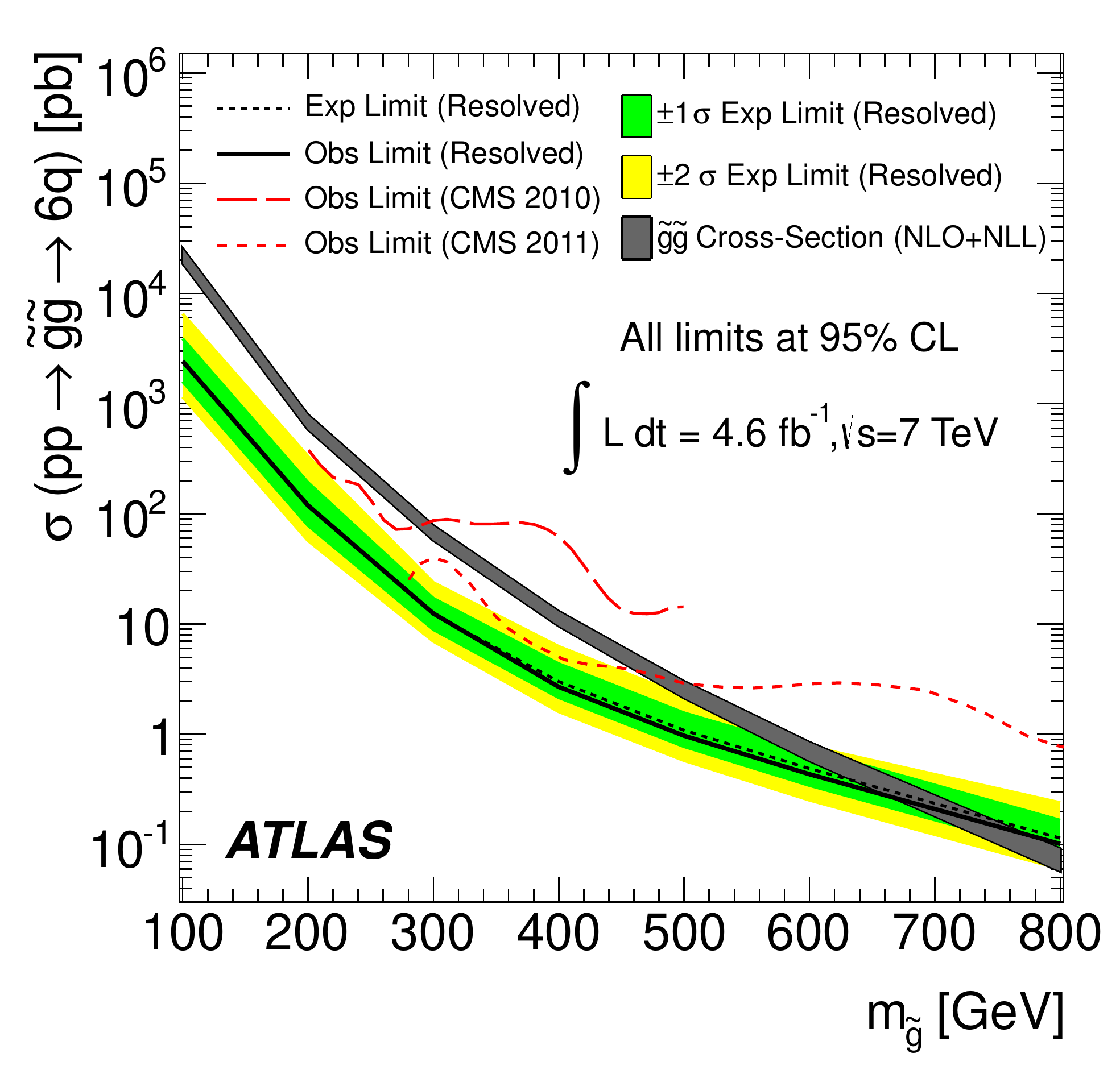}
    \caption{The expected and observed 95\% confidence limits are shown for the resolved analyses channel. The published CMS results using 35\invpb of 2010 data and using 5\invfb of 2011 data are shown for comparison.}

  \label{fig:LimitsResolved}
\end{figure*}

\begin{figure*}
  \centering
      \includegraphics[width=0.9\textwidth]{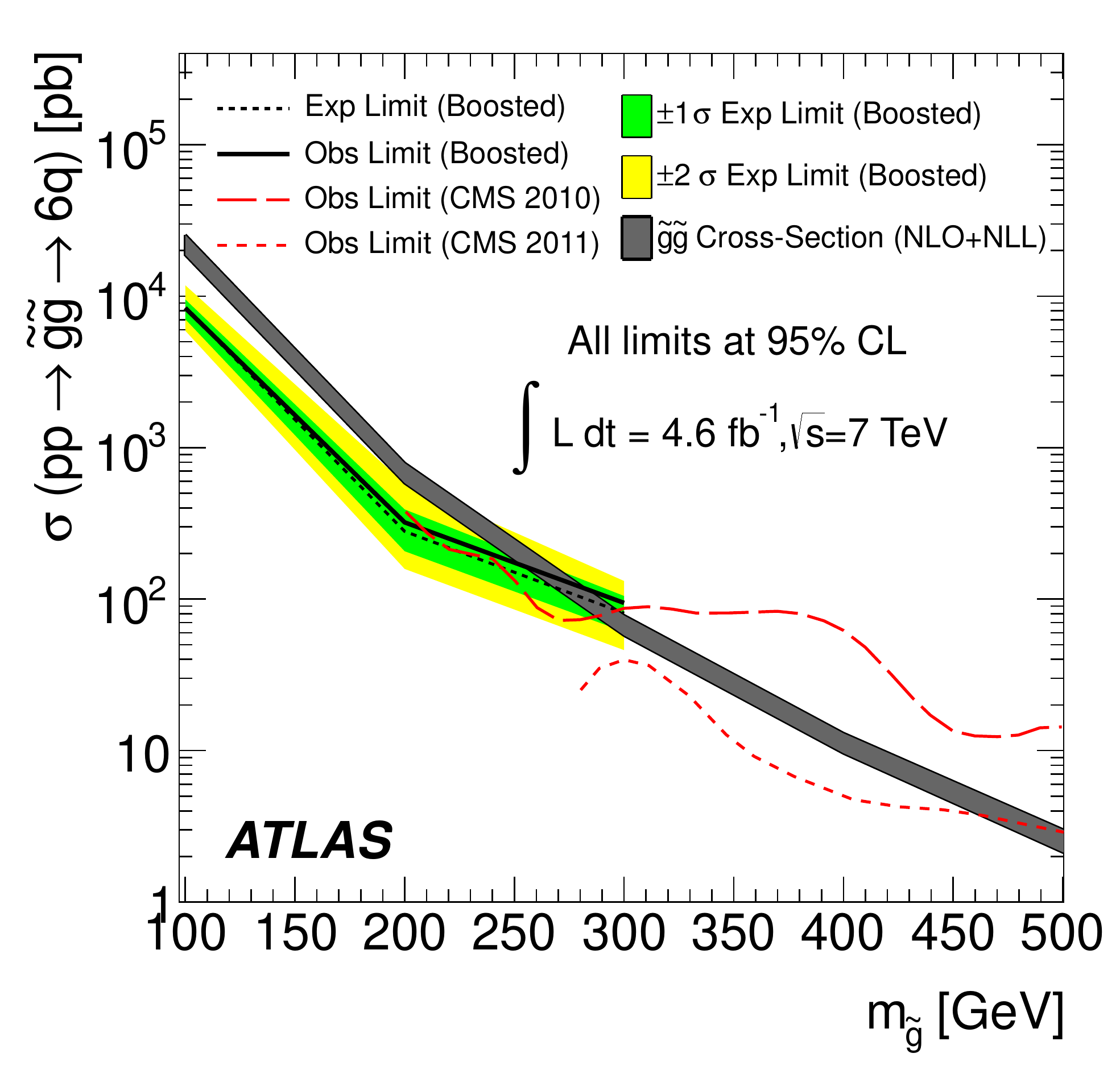}
    \caption{The expected and observed 95\% confidence limits are shown for the boosted analyses channel. The published CMS results using 35\invpb of 2010 data and using 5\invfb of 2011 data are shown for comparison.}

  \label{fig:LimitsBoosted}
\end{figure*}

Since no excess is observed in data in either analysis channel, a limit-setting procedure is performed. A profile likelihood ratio combining Poisson probabilities for signal and background is computed to determine the confidence level for consistency of the data with the signal-plus-background hypothesis (\CLsb). A similar calculation is performed for the background-only hypothesis only (\CLb). From the ratio of these two quantities, the confidence level for the presence of signal (\CLs) is determined~\cite{HistFitter}. Systematic uncertainties are treated via nuisance parameters assuming Gaussian distributions. The resulting expected and observed limits for each analysis channel are shown in figures~\ref{fig:LimitsResolved} and~\ref{fig:LimitsBoosted}. Mass limits are determined by comparing the observed and expected cross-section limits with the lower edge of the $\pm 1 \sigma$\ uncertainty band around the theoretical NLO+NLL cross-section prediction. This cross-section and the relevant acceptances for signal events to meet analysis requirements are summarized in \tabref{SignalProperties}. The boosted approach is sensitive to the low gluino mass region where gluinos may be produced with transverse momenta significantly greater than their mass. At the 95\% confidence level, this approach is able to exclude gluino masses $\mgluino<\ObsLimitBoosted$, as compared to an expected lower limit on the allowed gluino mass of \ExpLimitBoosted. Using the resolved approach, the observed lower limit on the allowed gluino mass is \ObsLimitResolved, whereas the expected limit is \ExpLimitResolved. It should be emphasized that the main systematic uncertainties on the background prediction are different for the two analyses, and the selected event samples are almost orthogonal to one another (less than 8\% overlap) in both the signal and the control regions of the two analyses. The results of the two analysis channels are therefore almost completely uncorrelated.

The resolved approach maintains a significant sensitivity even at large gluino masses, as expected. The sensitivity is still comparatively better than that of the boosted selection at low masses, despite the low mass region being the focus of the latter approach. This difference is primarily due to the high signal purity in the low-mass signal region for the resolved analysis, as well as the larger potential signal contamination of the background estimation for the boosted selection.

One must bear in mind that these limits are appropriate for the particular model that we have chosen in which the gluinos decay via off-shell squarks, and for the particular showering scheme that has been chosen. As discussed previously, since colour-flows are not well-understood in this final state and may be substantially different for other models, we do not include showering uncertainties in these results. Any differences in such modeling characteristics must be accounted for when reinterpreting these results.

\begin{table}[th]
\begin{center}
\begin{tabular}{l|c|c|c|c}

\hline  \hline

\multirow{2}{*}{Model (\mgluino)} & \multirow{2}{*}{$\sigma^{\textnormal{theory}}_{\textnormal{min}}$ [pb]} & \multirow{2}{*}{$\sigma^{\textnormal{theory}}_{\textnormal{max}}$ [pb]} & \multicolumn{2}{c}{Acceptance (\%)} \\
     & & & Resolved & Boosted \\
          \hline
    $100$~GeV   &  18700  &  25400  & 0.098  & 0.077 \\
    $200$~GeV   &  584    &  790    & 0.094 & 0.070 \\
    $300$~GeV   &  57.6   &  77.9   & 0.451  & 0.182 \\
    $400$~GeV   &  9.61   &  13.0   & 0.210  & --  \\
    $500$~GeV   &  2.13   &  3.01   & 0.565  & --  \\
    $600$~GeV   &  0.574  &  0.843  & 1.30   & --  \\
    $800$~GeV   &  0.0572 &  0.0913 & 5.73   & --  \\
    \hline \hline

\end{tabular}
    
    \caption{Cross-sections and acceptances for each of the signal samples used in the analysis. The trends in the acceptances are sometimes discontinuous due to the different signal regions that were chosen when optimising for different masses. } \label{tab:SignalProperties}

\end{center}
\end{table}


\clearpage
\section{Conclusions}
\label{sec:Conclusions}
The results of a search for pair production of heavy particles decaying into six-quark final states using two complementary analysis channels are reported. This search is carried out using an integrated luminosity of 4.6\invfb of \sqsseven \pp collisions at the LHC collected by the ATLAS detector. In one analysis channel, the number of events with at least six jets satisfying a particular \pT\ requirement is compared with expectations. In the second analysis channel, which is specifically intended to search for low-mass gluinos, a search is performed for highly boosted jets in which each gluino deposits its energy entirely within a single large radius  cone. In each analysis channel, results are observed to be fully consistent with the Standard Model. Using an RPV gluino-decay signal as a benchmark model, we set the most stringent limits on the model to date. For the resolved analysis channel, in the absence of a signal, 95\% exclusion limits are expected to exclude the region up to \ExpLimitResolved\ and are observed to exclude up to \ObsLimitResolved. For the boosted analysis channel, limits are expected up to \ExpLimitBoosted\ and are observed up to \ObsLimitBoosted.

%
%

\clearpage
\section{Acknowledgements}
\label{sec:Ack}

We thank CERN for the very successful operation of the LHC, as well as the
support staff from our institutions without whom ATLAS could not be
operated efficiently.

We acknowledge the support of ANPCyT, Argentina; YerPhI, Armenia; ARC,
Australia; BMWF and FWF, Austria; ANAS, Azerbaijan; SSTC, Belarus; CNPq and FAPESP,
Brazil; NSERC, NRC and CFI, Canada; CERN; CONICYT, Chile; CAS, MOST and NSFC,
China; COLCIENCIAS, Colombia; MSMT CR, MPO CR and VSC CR, Czech Republic;
DNRF, DNSRC and Lundbeck Foundation, Denmark; EPLANET, ERC and NSRF, European Union;
IN2P3-CNRS, CEA-DSM/IRFU, France; GNSF, Georgia; BMBF, DFG, HGF, MPG and AvH
Foundation, Germany; GSRT and NSRF, Greece; ISF, MINERVA, GIF, DIP and Benoziyo Center,
Israel; INFN, Italy; MEXT and JSPS, Japan; CNRST, Morocco; FOM and NWO,
Netherlands; BRF and RCN, Norway; MNiSW, Poland; GRICES and FCT, Portugal; MERYS
(MECTS), Romania; MES of Russia and ROSATOM, Russian Federation; JINR; MSTD,
Serbia; MSSR, Slovakia; ARRS and MVZT, Slovenia; DST/NRF, South Africa;
MICINN, Spain; SRC and Wallenberg Foundation, Sweden; SER, SNSF and Cantons of
Bern and Geneva, Switzerland; NSC, Taiwan; TAEK, Turkey; STFC, the Royal
Society and Leverhulme Trust, United Kingdom; DOE and NSF, United States of
America.

The crucial computing support from all WLCG partners is acknowledged
gratefully, in particular from CERN and the ATLAS Tier-1 facilities at
TRIUMF (Canada), NDGF (Denmark, Norway, Sweden), CC-IN2P3 (France),
KIT/GridKA (Germany), INFN-CNAF (Italy), NL-T1 (Netherlands), PIC (Spain),
ASGC (Taiwan), RAL (UK) and BNL (USA) and in the Tier-2 facilities
worldwide.


\clearpage


\begin{thebibliography}{10}

\bibitem{Miyazawa:1966}
H.~Miyazawa, {\em {Baryon Number Changing Currents}\/},
\href{http://dx.doi.org/10.1143/PTP.36.1266}{Prog. Theor. Phys. {\bf 36 (6)}
  (1966)  1266--1276}.

\bibitem{Ramond:1971gb}
P.~Ramond, {\em {Dual Theory for Free Fermions}\/},
\href{http://dx.doi.org/10.1103/PhysRevD.3.2415}{Phys. Rev. {\bf D 3} (1971)
  2415}.

\bibitem{Golfand:1971iw}
Y.~A. Gol'fand and E.~P. Likhtman, {\em {Extension of the Algebra of Poincare
  Group Generators and Violation of p Invariance}\/},  JETP Lett. {\bf 13}
  (1971)  323--326.
[Pisma Zh.Eksp.Teor.Fiz.13:452-455,1971].

\bibitem{Neveu:1971rx}
A.~Neveu and J.~H. Schwarz, {\em {Factorizable dual model of pions}\/},
\href{http://dx.doi.org/10.1016/0550-3213(71)90448-2}{Nucl. Phys. {\bf B 31}
  (1971)  86}.

\bibitem{Neveu:1971iv}
A.~Neveu and J.~H. Schwarz, {\em {Quark Model of Dual Pions}\/},
\href{http://dx.doi.org/10.1103/PhysRevD.4.1109}{Phys. Rev. {\bf D 4} (1971)
  1109}.

\bibitem{Gervais:1971ji}
J.~Gervais and B.~Sakita, {\em {Field theory interpretation of supergauges in
  dual models}\/},
\href{http://dx.doi.org/10.1016/0550-3213(71)90351-8}{Nucl. Phys. {\bf B 34}
  (1971)  632}.

\bibitem{Volkov:1973ix}
D.~V. Volkov and V.~P. Akulov, {\em {Is the Neutrino a Goldstone Particle?}\/},
\href{http://dx.doi.org/10.1016/0370-2693(73)90490-5}{Phys. Lett. {\bf B 46}
  (1973)  109}.

\bibitem{Wess:1973kz}
J.~Wess and B.~Zumino, {\em {A Lagrangian Model Invariant Under Supergauge
  Transformations}\/},
\href{http://dx.doi.org/10.1016/0370-2693(74)90578-4}{Phys. Lett. {\bf B 49}
  (1974)  52}.

\bibitem{Wess:1974tw}
J.~Wess and B.~Zumino, {\em {Supergauge Transformations in Four-Dimensions}\/},
\href{http://dx.doi.org/10.1016/0550-3213(74)90355-1}{Nucl. Phys. {\bf B 70}
  (1974)  39}.

\bibitem{Dimopoulos:1981zb}
S.~Dimopoulos and H.~Georgi, {\em {Softly Broken Supersymmetry and SU(5)}\/},
\href{http://dx.doi.org/10.1016/0550-3213(81)90522-8}{Nucl. Phys. {\bf B 193}
  (1981)  150}.

\bibitem{Witten:1981nf}
E.~Witten, {\em {Dynamical Breaking of Supersymmetry}\/},
\href{http://dx.doi.org/10.1016/0550-3213(81)90006-7}{Nucl. Phys. {\bf B 188}
  (1981)  513}.

\bibitem{Dine:1981za}
M.~Dine, W.~Fischler, and M.~Srednicki, {\em {Supersymmetric Technicolor}\/},
\href{http://dx.doi.org/10.1016/0550-3213(81)90582-4}{Nucl. Phys. {\bf B 189}
  (1981)  575--593}.

\bibitem{Dimopoulos:1981au}
S.~Dimopoulos and S.~Raby, {\em {Supercolor}\/},
\href{http://dx.doi.org/10.1016/0550-3213(81)90430-2}{Nucl. Phys. {\bf B 192}
  (1981)  353}.

\bibitem{Sakai:1981gr}
N.~Sakai, {\em {Naturalness in Supersymmetric Guts}\/},
\href{http://dx.doi.org/10.1007/BF01573998}{Zeit. Phys. {\bf C 11} (1981)
  153}.

\bibitem{Kaul:1981hi}
R.~Kaul and P.~Majumdar, {\em {Cancellation of quadratically divergent mass
  corrections in globally supersymmetric spontaneously broken gauge
  theories}\/},
\href{http://dx.doi.org/10.1016/0550-3213(82)90565-X}{Nucl. Phys. {\bf B 199}
  (1982)  36}.

\bibitem{Goldberg:1983nd}
H.~Goldberg, {\em {Constraint on the photino mass from cosmology}\/},
\href{http://dx.doi.org/10.1103/PhysRevLett.50.1419}{Phys. Rev. Lett. {\bf 50}
  (1983)  1419}.

\bibitem{Ellis:1983ew}
J.~Ellis, J.~Hagelin, D.~Nanopoulos, K.~Olive, and M.~Srednicki, {\em
  {Supersymmetric relics from the big bang}\/},
\href{http://dx.doi.org/10.1016/0550-3213(84)90461-9}{Nucl. Phys. {\bf B 238}
  (1984)  453}.

\bibitem{SUSYZeroLep2011}
{ATLAS} Collaboration, {\em {Search for squarks and gluinos using final states
  with jets and missing transverse momentum with the ATLAS detector in
  $\sqrt{s} = 7$~TeV proton-proton collisions}\/},
  \href{http://dx.doi.org/10.1016/j.physletb.2012.02.051}{Phys. Lett. {\bf B
  710} (2012)  67},
\href{http://arxiv.org/abs/1109.6572}{{\tt arXiv:1109.6572 [hep-ex]}}.

\bibitem{SUSYOneLep2011}
{ATLAS} Collaboration, {\em {Search for supersymmetry in final states with
  jets, missing transverse momentum and one isolated lepton in $\sqrt{s} =
  7$~TeV $pp$ collisions using 1 $fb^{-1}$ of ATLAS data}\/},
  \href{http://dx.doi.org/10.1103/PhysRevD.85.012006}{Phys. Rev. {\bf D 85}
  (2012)  012006},
\href{http://arxiv.org/abs/1109.6606}{{\tt arXiv:1109.6606 [hep-ex]}}.

\bibitem{SUSYTwoLep2011}
{ATLAS} Collaboration, {\em {Searches for supersymmetry with the ATLAS detector
  using final states with two leptons and missing transverse momentum in
  $\sqrt{s} = 7$~TeV proton-proton collisions}\/},
  \href{http://dx.doi.org/10.1016/j.physletb.2012.01.076}{Phys. Lett. {\bf B
  709} (2012)  137},
\href{http://arxiv.org/abs/1110.6189}{{\tt arXiv:1110.6189 [hep-ex]}}.

\bibitem{SUSYJetMult2011}
{ATLAS} Collaboration, {\em {Search for new phenomena in final states with
  large jet multiplicities and missing transverse momentum using $\sqrt{s} =
  7$~TeV $pp$ collisions with the ATLAS detector}\/},
  \href{http://dx.doi.org/10.1007/JHEP11(2011)099}{JHEP {\bf 1111} (2011)
  099},
\href{http://arxiv.org/abs/1110.2299}{{\tt arXiv:1110.2299 [hep-ex]}}.

\bibitem{Aaltonen:2011sg}
{CDF} Collaboration, {\em {First Search for Multijet Resonances in $\sqrt{s} =
  1.96$ TeV $ p\bar{p}$ Collisions}\/},
  \href{http://dx.doi.org/10.1103/PhysRevLett.107.042001}{Phys. Rev. Lett. {\bf
  107} (2011)  042001},
\href{http://arxiv.org/abs/1105.2815}{{\tt arXiv:1105.2815 [hep-ex]}}.

\bibitem{Chatrchyan:2011cj}
{CMS} Collaboration, {\em {Search for Three-Jet Resonances in $pp$ Collisions
  at $\sqrt{s} = 7$~TeV}\/},
  \href{http://dx.doi.org/10.1103/PhysRevLett.107.101801}{Phys. Rev. Lett. {\bf
  107} (2011)  101801},
\href{http://arxiv.org/abs/1107.3084}{{\tt arXiv:1107.3084 [hep-ex]}}.

\bibitem{CMSRPVGluino2011}
{CMS} Collaboration, {\em {Search for three-jet resonances in $pp$ collisions
  at $\sqrt{s} = 7$~TeV}\/},
\href{http://arxiv.org/abs/1208.2931}{{\tt arXiv:1208.2931 [hep-ex]}}.

\bibitem{detPaper}
{ATLAS Collaboration}, {\em {The ATLAS Experiment at the CERN Large Hadron
  Collider}\/},
\href{http://dx.doi.org/10.1088/1748-0221/3/08/S08003}{JINST {\bf 3} (2008)
  S08003}.

\bibitem{PerfWithData2010}
{ATLAS Collaboration}, {\em {Performance of the ATLAS Detector using First
  Collision Data}\/},  \href{http://dx.doi.org/10.1007/JHEP09(2010)056}{JHEP
  {\bf 09} (2010)  056},
\href{http://arxiv.org/abs/hep-ex/1005.5254}{{\tt arXiv:hep-ex/1005.5254
  [hep-ex]}}.

\bibitem{LArReadiness}
{ATLAS Collaboration}, {\em {Readiness of the {ATLAS} liquid argon calorimeter
  for {LHC} collisions}\/},
  \href{http://dx.doi.org/10.1140/epjc/s10052-010-1354-y}{Eur. Phys. J. {\bf C
  70} (2010)  723}, \href{http://arxiv.org/abs/0912.2642}{{\tt arXiv:0912.2642
  [physics.ins-det]}}.

\bibitem{TileReadiness}
{ATLAS} Collaboration, {\em {Readiness of the {ATLAS} Tile calorimeter for
  {LHC} collisions}\/},
  \href{http://dx.doi.org/10.1140/epjc/s10052-010-1508-y}{Eur. Phys. J. {\bf C
  70} (2010)  1193}, \href{http://arxiv.org/abs/1007.5423}{{\tt arXiv:1007.5423
  [physics.ins-det]}}.

\bibitem{TriggerPerf2010}
{ATLAS Collaboration}, {\em {Performance of the ATLAS Trigger System in
  2010}\/},  \href{http://dx.doi.org/10.1140/epjc/s10052-011-1849-1}{Eur. Phys.
  J. {\bf C 72} (2012)  1849}, \href{http://arxiv.org/abs/1110.1530}{{\tt
  arXiv:1110.1530 [hep-ex]}}.

\bibitem{jespaper2010}
{ATLAS} Collaboration, {\em {Jet energy measurement with the ATLAS detector in
  proton-proton collisions at $\sqrt{s} = 7$~TeV}\/},  Submitted to EPJC (2011)
   , \href{http://arxiv.org/abs/1112.6426}{{\tt arXiv:1112.6426 [hep-ex]}}.

\bibitem{ctb2004electronseoverp}
E.~Abat et al., {\em {Combined performance studies for electrons at the 2004
  {ATLAS} combined test-beam}\/},
\href{http://dx.doi.org/10.1088/1748-0221/5/11/P11006}{JINST {\bf 5} (2010)
  P11006}.

\bibitem{ctb2004electrons}
M.~Aharrouche et al., {\em {Measurement of the response of the {ATLAS} liquid
  argon barrel calorimeter to electrons at the 2004 combined test-beam}\/},
\href{http://dx.doi.org/10.1016/j.nima.2009.12.055}{Nucl. Instrum. Meth. {\bf A
  614} (2010)  400}.

\bibitem{LArTB02uniformity}
J.~Colas et al., {\em {Response uniformity of the {ATLAS} liquid argon
  electromagnetic calorimeter}\/},
  \href{http://dx.doi.org/10.1016/j.nima.2007.08.157}{Nucl. Instrum. Meth. {\bf
  A 582} (2007)  429},
\href{http://arxiv.org/abs/0709.1094}{{\tt arXiv:0709.1094 [physics.ins-det]}}.

\bibitem{LArTB02linearity}
M.~Aharrouche et al., {\em {Energy linearity and resolution of the {ATLAS}
  electromagnetic barrel calorimeter in an electron test-beam}\/},
  \href{http://dx.doi.org/10.1016/j.nima.2006.07.053}{Nucl. Instrum. Meth. {\bf
  A 568} (2006)  601}.

\bibitem{LArTB02muons}
M.~Aharrouche et al., {\em {Study of the response of {ATLAS} electromagnetic
  liquid argon calorimeters to muons}\/},
\href{http://dx.doi.org/10.1016/j.nima.2009.05.021}{Nucl. Instrum. Meth. {\bf A
  606} (2009)  419}.

\bibitem{Tile2002}
P.~Adragna et al., {\em {Testbeam studies of production modules of the {ATLAS}
  {T}ile calorimeter}\/},
\href{http://dx.doi.org/10.1016/j.nima.2009.04.009}{Nucl. Instrum. Meth. {\bf A
  606} (2009)  362}.

\bibitem{Tile2002pionproton}
P.~Adragna et al., {\em {Measurement of pion and proton response and
  longitudinal shower profiles up to 20 nuclear interaction lengths with the
  ATLAS Tile calorimeter}\/},
\href{http://dx.doi.org/10.1016/j.nima.2010.01.037}{Nucl. Instrum. Meth. {\bf A
  615} (2010)  158}.

\bibitem{CTB2004topology}
E.~Abat et al., {\em {Response and shower topology of 2 to 180~GeV pions
  measured with the ATLAS barrel calorimeter at the CERN test-beam and
  comparison to Monte Carlo simulations}\/},
  \href{http://cdsweb.cern.ch/record/1263861}{ATL-CAL-PUB-2010-001}.
  \url{http://cdsweb.cern.ch/record/1263861}.

\bibitem{CTB04pion}
E.~Abat et al., {\em {Study of energy response and resolution of the ATLAS
  barrel calorimeter to hadrons of energies from 20~GeV to 35~GeV}\/},
\href{http://dx.doi.org/10.1016/j.nima.2010.04.054}{Nucl. Instrum. Meth. {\bf A
  621} (2010)  134}.

\bibitem{CTB2004vlepion}
E.~Abat et al., {\em Study of the response of the {ATLAS} central calorimeter
  to pions of energies from 3 to 9 {G}eV\/},
  \href{http://dx.doi.org/10.1016/j.nima.2009.05.158}{Nucl. Instrum. Meth. {\bf
  A 607} (2009)  372}.

\bibitem{EndcapTBelectronPion2002}
C.~Cojocaru et al., {\em Hadronic calibration of the {ATLAS} liquid argon
  end-cap calorimeter in the pseudorapidity region $1.6<|\eta|<1.8$ in beam
  tests\/},  \href{http://dx.doi.org/10.1016/j.nima.2004.05.133}{Nucl. Instrum.
  Meth. {\bf A 531} (2004)  481}.

\bibitem{Pinfold:2008zzb}
J.~Pinfold et al., {\em {Performance of the {ATLAS} liquid argon endcap
  calorimeter in the pseudorapidity region $2.5 < |\eta| < 4.0$ in beam
  tests}\/},
\href{http://dx.doi.org/10.1016/j.nima.2008.05.033}{Nucl. Instrum. Meth. {\bf A
  593} (2008)  324}.

\bibitem{Kiryunin:2006cm}
A.~E. Kiryunin, H.~Oberlack, D.~Salihagic, P.~Schacht, and P.~Strizenec, {\em
  {{GEANT4} physics evaluation with testbeam data of the {ATLAS} hadronic
  end-cap calorimeter}\/},
\href{http://dx.doi.org/10.1016/j.nima.2005.12.237}{Nucl. Instrum. Meth. {\bf A
  560} (2006)  278}.

\bibitem{Atlaselectronpaper}
{ATLAS} Collaboration, {\em {Electron performance measurements with the ATLAS
  detector using the 2010 LHC proton-proton collision data}\/},
  \href{http://dx.doi.org/10.1140/epjc/s10052-012-1909-1}{Eur. Phys. J. {\bf C
  72} (2012)  1909}, \href{http://arxiv.org/abs/1110.3174}{{\tt arXiv:1110.3174
  [hep-ex]}}.

\bibitem{Cacciari200657}
M.~Cacciari and G.~P. Salam, {\em {Dispelling the $N^{3}$ myth for the $k_t$
  jet-finder}\/},
  \href{http://dx.doi.org/10.1016/j.physletb.2006.08.037}{Phys. Lett. {\bf B
  641} (2006)  57}, \href{http://arxiv.org/abs/hep-ph/0512210}{{\tt
  arXiv:hep-ph/0512210 [hep-ph]}}.

\bibitem{Cacciari:2008gp}
M.~Cacciari, G.~P. Salam, and G.~Soyez, {\em {The anti-$k_{t}$ jet clustering
  algorithm}\/},  \href{http://dx.doi.org/10.1088/1126-6708/2008/04/063}{JHEP
  {\bf 04} (2008)  063},
\href{http://arxiv.org/abs/0802.1189}{{\tt arXiv:0802.1189 [hep-ph]}}.

\bibitem{TopoClusters}
W.~Lampl et al., {\em Calorimeter clustering algorithms: description and
  performance\/},
  \href{http://cdsweb.cern.ch/record/1099735}{ATL-LARG-PUB-2008-002}.
  \url{http://cdsweb.cern.ch/record/1099735}.

\bibitem{Aad:2011dr}
{ATLAS} Collaboration, {\em {Luminosity determination in p-p collisions at
  $\sqrt{s}=7$~TeV using the {ATLAS} detector at the LHC}\/},
  \href{http://dx.doi.org/10.1140/epjc/s10052-011-1630-5}{Eur. Phys. J. {\bf C
  71} (2011)  1630}, \href{http://arxiv.org/abs/1101.2185}{{\tt arXiv:1101.2185
  [hep-ex]}}.

\bibitem{ATLAS-CONF-2011-116}
{ATLAS} Collaboration, {\em {Luminosity Determination in pp Collisions at
  $\sqrt{s}=7$~TeV using the ATLAS Detector in 2011}\/},
  \href{http://cdsweb.cern.ch/record/1376384}{ATLAS-CONF-2011-116}.
  \url{http://cdsweb.cern.ch/record/1376384}.

\bibitem{Alwall:2007st}
J.~Alwall et al., {\em {MadGraph/MadEvent v4: The New Web Generation}\/},
  \href{http://dx.doi.org/10.1088/1126-6708/2007/09/028}{JHEP {\bf 09} (2007)
  028},
\href{http://arxiv.org/abs/0706.2334}{{\tt arXiv:0706.2334 [hep-ph]}}.

\bibitem{Alwall:2011uj}
J.~Alwall, M.~Herquet, F.~Maltoni, O.~Mattelaer, and T.~Stelzer, {\em {MadGraph
  5 : Going Beyond}\/},  \href{http://dx.doi.org/10.1007/JHEP06(2011)128}{JHEP
  {\bf 1106} (2011)  128},
\href{http://arxiv.org/abs/1106.0522}{{\tt arXiv:1106.0522 [hep-ph]}}.

\bibitem{Fuks:2012im}
B.~Fuks, {\em {Beyond the Minimal Supersymmetric Standard Model: from theory to
  phenomenology}\/},  \href{http://dx.doi.org/10.1142/S0217751X12300074}{Int.
  J. Mod. Phys. {\bf A 27} (2012)  1230007},
\href{http://arxiv.org/abs/1202.4769}{{\tt arXiv:1202.4769 [hep-ph]}}.

\bibitem{pythia8}
T.~Sjostrand, S.~Mrenna, and P.~Z. Skands, {\em {A Brief Introduction to PYTHIA
  8.1}\/},  \href{http://dx.doi.org/10.1016/j.cpc.2008.01.036}{Comput. Phys.
  Commun. {\bf 178} (2008)  852--867},
\href{http://arxiv.org/abs/0710.3820}{{\tt arXiv:0710.3820 [hep-ph]}}.

\bibitem{Beenakker:1996ch}
W.~Beenakker, R.~Hopker, M.~Spira, and P.~Zerwas, {\em {Squark and gluino
  production at hadron colliders}\/},
  \href{http://dx.doi.org/10.1016/S0550-3213(97)00084-9}{Nucl. Phys. {\bf B
  492} (1997)  51},
\href{http://arxiv.org/abs/hep-ph/9610490}{{\tt arXiv:hep-ph/9610490
  [hep-ph]}}.

\bibitem{Kulesza:2008jb}
A.~Kulesza and L.~Motyka, {\em {Threshold resummation for squark-antisquark and
  gluino-pair production at the LHC}\/},
  \href{http://dx.doi.org/10.1103/PhysRevLett.102.111802}{Phys. Rev. Lett. {\bf
  102} (2009)  111802},
\href{http://arxiv.org/abs/0807.2405}{{\tt arXiv:0807.2405 [hep-ph]}}.

\bibitem{Kulesza:2009kq}
A.~Kulesza and L.~Motyka, {\em {Soft gluon resummation for the production of
  gluino-gluino and squark-antisquark pairs at the LHC}\/},
  \href{http://dx.doi.org/10.1103/PhysRevD.80.095004}{Phys. Rev. {\bf D 80}
  (2009)  095004},
\href{http://arxiv.org/abs/0905.4749}{{\tt arXiv:0905.4749 [hep-ph]}}.

\bibitem{Beenakker:2009ha}
W.~Beenakker, S.~Brensing, M.~Kramer, A.~Kulesza, E.~Laenen, et al., {\em
  {Soft-gluon resummation for squark and gluino hadroproduction}\/},
  \href{http://dx.doi.org/10.1088/1126-6708/2009/12/041}{JHEP {\bf 0912} (2009)
   041},
\href{http://arxiv.org/abs/0909.4418}{{\tt arXiv:0909.4418 [hep-ph]}}.

\bibitem{Beenakker:2011fu}
W.~Beenakker, S.~Brensing, M.~Kramer, A.~Kulesza, E.~Laenen, et al., {\em
  {Squark and gluino hadroproduction}\/},
  \href{http://dx.doi.org/10.1142/S0217751X11053560}{Int. J. Mod. Phys. {\bf A
  26} (2011)  2637},
\href{http://arxiv.org/abs/1105.1110}{{\tt arXiv:1105.1110 [hep-ph]}}.

\bibitem{Kramer:2012bx}
M.~Kramer, A.~Kulesza, R.~van~der Leeuw, M.~Mangano, S.~Padhi, et al., {\em
  {Supersymmetry production cross sections in pp collisions at $\sqrt{s} =
  7$~TeV}\/},
\href{http://arxiv.org/abs/1206.2892}{{\tt arXiv:1206.2892 [hep-ph]}}.

\bibitem{pythia}
T.~Sjostrand, S.~Mrenna, and P.~Z. Skands, {\em {PYTHIA 6.4 physics and
  manual}\/},  \href{http://dx.doi.org/10.1088/1126-6708/2006/05/026}{JHEP {\bf
  0605} (2006)  026},
\href{http://arxiv.org/abs/0603175}{{\tt arXiv:0603175 [hep-ph]}}.

\bibitem{MC11}
{ATLAS} Collaboration, {\em New ATLAS event generator tunes to 2010 data\/},
  \href{http://cdsweb.cern.ch/record/1345343}{ATL-PHYS-PUB-2011-008}.
  \url{http://cdsweb.cern.ch/record/1345343}.

\bibitem{MC11c}
{ATLAS} Collaboration, {\em ATLAS tunes for Pythia6 and Pythia8 for MC11\/},
  \href{http://cdsweb.cern.ch/record/1363300}{ATLAS-PHYS-PUB-2011-009}.
  \url{http://cdsweb.cern.ch/record/1363300}.

\bibitem{Nason:2004rx}
P.~Nason, {\em {A New method for combining NLO QCD with shower Monte Carlo
  algorithms}\/},  \href{http://dx.doi.org/10.1088/1126-6708/2004/11/040}{JHEP
  {\bf 0411} (2004)  040}, \href{http://arxiv.org/abs/hep-ph/0409146}{{\tt
  arXiv:hep-ph/0409146 [hep-ph]}}.

\bibitem{Frixione:2007vw}
S.~Frixione, P.~Nason, and C.~Oleari, {\em {Matching NLO QCD computations with
  Parton Shower simulations: the POWHEG method}\/},
  \href{http://dx.doi.org/10.1088/1126-6708/2007/11/070}{JHEP {\bf 0711} (2007)
   070}, \href{http://arxiv.org/abs/0709.2092}{{\tt arXiv:0709.2092 [hep-ph]}}.

\bibitem{groomedCONF2011}
{ATLAS} Collaboration, {\em {Performance of large-$R$ jets and jet substructure
  reconstruction with the ATLAS detector}\/},
  \href{http://cdsweb.cern.ch/record/1459530}{ATLAS-CONF-2011-065}.
  \url{http://cdsweb.cern.ch/record/1459530}.

\bibitem{simulation}
{ATLAS} Collaboration, {\em {The {ATLAS} simulation infrastructure}\/},
  \href{http://dx.doi.org/10.1140/epjc/s10052-010-1429-9}{Eur. Phys. J. {\bf C
  70} (2010)  823}, \href{http://arxiv.org/abs/1005.4568}{{\tt arXiv:1005.4568
  [physics.ins-det]}}.

\bibitem{Geant4}
{GEANT4} Collaboration, S.~Agostinelli et al., {\em {GEANT4: A simulation
  toolkit}\/},
\href{http://dx.doi.org/10.1016/S0168-9002(03)01368-8}{Nucl. Instrum. Meth.
  {\bf A 506} (2003)  250}.

\bibitem{PDF-CTEQ}
J.~Pumplin et al., {\em {New generation of parton distributions with
  uncertainties from global QCD analysis}\/},
  \href{http://dx.doi.org/10.1088/1126-6708/2002/07/012}{JHEP {\bf 07} (2002)
  012},
\href{http://arxiv.org/abs/0201195}{{\tt arXiv:0201195 [hep-ph]}}.

\bibitem{cteq6l1}
P.~M. Nadolsky et al., {\em Implications of CTEQ global analysis for collider
  observables\/},  \href{http://dx.doi.org/10.1103/PhysRevD.78.013004}{Phys.
  Rev. {\bf D 78} (2008)  013004},
\href{http://arxiv.org/abs/0802.0007}{{\tt arXiv:0802.0007 [hep-ph]}}.

\bibitem{Whalley:2005nh}
M.~Whalley, D.~Bourilkov, and R.~Group, {\em {The Les Houches accord PDFs
  (LHAPDF) and LHAGLUE}\/},
\href{http://arxiv.org/abs/hep-ph/0508110}{{\tt arXiv:hep-ph/0508110
  [hep-ph]}}.

\bibitem{Ball:2008by}
{NNPDF} Collaboration, R.~D. Ball et al., {\em {A Determination of parton
  distributions with faithful uncertainty estimation}\/},
  \href{http://dx.doi.org/10.1016/j.nuclphysb.2008.09.037,
  10.1016/j.nuclphysb.2009.02.027}{Nucl.Phys. {\bf B 809} (2009)  1},
\href{http://arxiv.org/abs/0808.1231}{{\tt arXiv:0808.1231 [hep-ph]}}.

\bibitem{Ball:2010de}
R.~D. Ball, L.~Del~Debbio, S.~Forte, A.~Guffanti, J.~I. Latorre, et al., {\em
  {A first unbiased global NLO determination of parton distributions and their
  uncertainties}\/},
  \href{http://dx.doi.org/10.1016/j.nuclphysb.2010.05.008}{Nucl.Phys. {\bf B
  838} (2010)  136},
\href{http://arxiv.org/abs/1002.4407}{{\tt arXiv:1002.4407 [hep-ph]}}.

\bibitem{PDF-MRST}
A.~Sherstnev and R.~S. Thorne, {\em {Parton distributions for LO
  generators}\/},  \href{http://dx.doi.org/10.1140/epjc/s10052-008-0610-x}{Eur.
  Phys. J. {\bf C 55} (2008)  553},
\href{http://arxiv.org/abs/0711.2473}{{\tt arXiv:0711.2473 [hep-ph]}}.

\bibitem{Martin:2009iq}
A.~Martin, W.~Stirling, R.~Thorne, and G.~Watt, {\em {Parton distributions for
  the LHC}\/},  \href{http://dx.doi.org/10.1140/epjc/s10052-009-1072-5}{Eur.
  Phys. J. {\bf C 63} (2009)  189},
\href{http://arxiv.org/abs/0901.0002}{{\tt arXiv:0901.0002 [hep-ph]}}.

\bibitem{Desai:2011su}
N.~Desai and P.~Z. Skands, {\em {Supersymmetry and Generic BSM Models in PYTHIA
  8}\/},
\href{http://arxiv.org/abs/1109.5852}{{\tt arXiv:1109.5852 [hep-ph]}}.

\bibitem{Thaler:2010tr}
J.~Thaler and K.~Van~Tilburg, {\em {Identifying Boosted Objects with
  $N$-subjettiness}\/},  \href{http://dx.doi.org/10.1007/JHEP03(2011)015}{JHEP
  {\bf 1103} (2011)  015}, \href{http://arxiv.org/abs/1011.2268}{{\tt
  arXiv:1011.2268 [hep-ph]}}.

\bibitem{Thaler:2011gf}
J.~Thaler and K.~Van~Tilburg, {\em {Maximizing Boosted Top Identification by
  Minimizing $N$-subjettiness (2011)}\/},
\href{http://arxiv.org/abs/1108.2701}{{\tt arXiv:1108.2701 [hep-ph]}}.

\bibitem{Catani1993}
S.~Catani, Y.~L. Dokshitzer, M.~Seymour, and B.~Webber, {\em {Longitudinally
  invariant $K_t$ clustering algorithms for hadron hadron collisions}\/},
\href{http://dx.doi.org/10.1016/0550-3213(93)90166-M}{Nucl. Phys. {\bf B 406}
  (1993)  187}.

\bibitem{Krohn2010}
D.~Krohn, J.~Thaler, and L.-T. Wang, {\em {Jet trimming}\/},
  \href{http://dx.doi.org/10.1007/JHEP02(2010)084}{JHEP {\bf 2010} (2010)  20},
  \href{http://arxiv.org/abs/0912.1342}{{\tt arXiv:0912.1342 [hep-ph]}}.

\bibitem{Ellis1993}
S.~D. Ellis and D.~E. Soper, {\em {Successive combination jet algorithm for
  hadron collisions}\/},
  \href{http://dx.doi.org/10.1103/PhysRevD.48.3160}{Phys. Rev. {\bf D 48}
  (1993)  3160},
\href{http://arxiv.org/abs/hep-ph/9305266}{{\tt arXiv:hep-ph/9305266
  [hep-ph]}}.

\bibitem{groomedJetPileup2011}
{ATLAS} Collaboration, {\em {Studies of the impact and mitigation of pile-up on
  large radius and groomed jets in ATLAS at $\sqrt{s}=7$~TeV}\/},
  \href{http://cdsweb.cern.ch/record/1459531}{ATLAS-CONF-2011-066}.
  \url{http://cdsweb.cern.ch/record/1459531}.

\bibitem{BOOST2011}
A.~Altheimer et al., {\em {Jet Substructure at the Tevatron and LHC: New
  results, new tools, new benchmarks}\/},
  \href{http://dx.doi.org/10.1088/0954-3899/39/6/063001}{J. Phys. {\bf G39}
  (2012)  063001},
\href{http://arxiv.org/abs/1201.0008}{{\tt arXiv:1201.0008 [hep-ph]}}.

\bibitem{HistFitter}
A.~L. Read, {\em {Presentation of search results: The CL(s) technique}\/},
\href{http://dx.doi.org/10.1088/0954-3899/28/10/313}{J. Phys. {\bf G28} (2002)
  2693--2704}.

\end{thebibliography}

\providecommand{\href}[2]{#2}\begingroup\raggedright\endgroup

\clearpage
\onecolumn
\begin{flushleft}
{\Large The ATLAS Collaboration}

\bigskip

G.~Aad$^{\rm 48}$,
T.~Abajyan$^{\rm 21}$,
B.~Abbott$^{\rm 111}$,
J.~Abdallah$^{\rm 12}$,
S.~Abdel~Khalek$^{\rm 115}$,
A.A.~Abdelalim$^{\rm 49}$,
O.~Abdinov$^{\rm 11}$,
R.~Aben$^{\rm 105}$,
B.~Abi$^{\rm 112}$,
M.~Abolins$^{\rm 88}$
% ...
\bigskip
\\
% ...
$^{11}$ Institute of Physics, Azerbaijan Academy of Sciences, Baku, Azerbaijan\\
$^{12}$ Institut de F{\'\i}sica d'Altes Energies and Departament de F{\'\i}sica de la Universitat Aut{\`o}noma de Barcelona and ICREA, Barcelona, Spain\\
% ...
$^{21}$ Physikalisches Institut, University of Bonn, Bonn, Germany\\
% ...
$^{48}$ Fakult{\"a}t f{\"u}r Mathematik und Physik, Albert-Ludwigs-Universit{\"a}t, Freiburg, Germany\\
$^{49}$ Section de Physique, Universit{\'e} de Gen{\`e}ve, Geneva, Switzerland\\
% ...
$^{88}$ Department of Physics and Astronomy, Michigan State University, East Lansing MI, United States of America\\
% ...
$^{105}$ Nikhef National Institute for Subatomic Physics and University of Amsterdam, Amsterdam, Netherlands\\
% ...
$^{111}$ Homer L. Dodge Department of Physics and Astronomy, University of Oklahoma, Norman OK, United States of America\\
$^{112}$ Department of Physics, Oklahoma State University, Stillwater OK, United States of America\\
% ...
$^{115}$ LAL, Universit{\'e} Paris-Sud and CNRS/IN2P3, Orsay, France\\
% ...
\end{flushleft}


\end{document}